\documentclass{aa}  

\usepackage{graphicx}
\usepackage{inputenc}
\graphicspath{../figures}
\usepackage{txfonts}
\def\ergs{ergs/cm${^2}$/s}
\def\relxill{{\it relxill}}
\def\relxilllp{{\it relxilllp}}
\def\nh{N$_{\rm H}$}
\def\afe{A$_{\rm Fe}$}
\def\logxi{$\log \xi$}
\def\rg{R$_{\rm g}$}
\def\rf{R$_{\rm f}$}
\def\cstat{cstat}
\def\zxipcf{{\it zxipcf}}

\begin{document}


   \title{Inferring black hole spins and probing accretion/ejection flows in AGNs with the Athena X-ray Integral Field Unit}

   \subtitle{}

   \author{Didier Barret
          \inst{1}  and
          Massimo Cappi\inst{2}}

   \institute{Universit\'e de Toulouse; CNRS;  Institut de Recherche en Astrophysique et Plan\'etologie; 9 Avenue du colonel Roche, BP 44346, F-31028 Toulouse cedex 4, France \\
   \and
  INAF-Osservatorio di astrofisica e scienza dello spazio di Bologna (OAS), Via Piero Gobetti 93/3, I-40129 Bologna, Italy \\
   \\
              \email{dbarret@irap.omp.eu}
             \thanks{Send questions \& comments to dbarret@irap.omp.eu}
               }
         
   \date{Received May 2nd, 2019; accepted June 3rd, 2019}

 
  \abstract
 {Active Galactic Nuclei (AGN) display complex X-ray spectra which exhibit a variety of emission and absorption features, that are commonly interpreted as a combination of i) a relativistically smeared reflection component, resulting from the irradiation of an accretion disk by a compact hard X-ray source, ii) one or several warm/ionized absorption components produced by AGN-driven outflows crossing our line of sight, and iii) a non relativistic reflection component produced by more distant material. Disentangling these components via detailed model fitting can thus be used to constrain the black hole spin, the geometry and characteristics of the accretion flow, as well as of the outflows and surroundings of the black hole.}
{We investigate how a high throughput high resolution X-ray spectrometer, such as the Athena X-ray Integral Field Unit (X-IFU) can be used to this aim, using the state of the art reflection model  \relxill\ in a lamp post geometrical configuration.}
{We simulate a representative sample of AGN spectra, including all necessary model complexities, as well as a range of model parameters going from standard to more extreme values, and considered X-ray fluxes that are representative of known AGN and Quasars (QSOs) populations. We also present a method to estimate the systematic errors related to the uncertainties in the calibration of the X-IFU.}
 {In a conservative setting, in which the reflection component is computed self consistently by the \relxill\ model from the pre-set geometry and no iron over abundance, the mean errors on the spin and height of the irradiating source are $<0.05$ and $\sim 0.2$ \rg (in units of gravitational radius).  Similarly the absorber parameters (column density, ionization parameter, covering factor and velocity) are measured to an accuracy typically less than $\sim 5$\% over their allowed range of variations. Extending the simulations to include blue shifted ultra fast outflows, we show that X-IFU could measure their velocity with statistical errors $< 1$\%, even for high redshift objects (e.g. at redshifts $\sim 2.5$). }
 {The simulations presented here demonstrate the potential of the X-IFU to understand how black holes are powered and how they shape their host galaxies. The accuracy to recover the physical model parameters encoded in their X-ray emission is reached thanks to the unique capability of X-IFU to separate and constrain, narrow and broad, emission and absorption components.}
\titlerunning{High resolution X-ray spectroscopy of accreting supermassive black holes}
\authorrunning{Didier Barret \& Massimo Cappi}
   \maketitle
%
\section{Introduction}

One flagship of AGN studies resides in the unique opportunity X-rays offer to directly probe the innermost regions around AGN central black holes. 
In fact, AGN spectral properties exhibiting numerous emission and absorption lines in X-rays, combined with the observed fast variability of both continuum and lines, provide unique tools to measure the velocities, the ionization states, the time variations and the geometries of the accretion/ejection flows surrounding supermassive black holes (SMBHs) \citep{done_2010_lessons_arxiv, fabian_2016_reverb,  Kaastra_2017_AN,  reynolds_2019_nat}.

With the advent of high throughput and high spectral resolution of the imaging and grating spectrometers on-board XMM-Newton and Chandra, the field has seen the flourishing of a number of detailed spectral and timing studies addressing two broad topics: i) the SMBH-accretion disk systems via the relativistic reflection component from the accretion disk and the inferred SMBH spins (e.g. \citet{tanaka_1995,nandra_1997,brenneman_2006_apj,fabian_2009_nature, zoghbi_2010_firstlag, demarco_2013_mnras, brenneman_2013, fabian_2017_AN, kara_2016_AN,garcia_2019_wpusds,zoghbi_2019_wpusds}), and ii) the study of AGN-driven outflows (from low velocity warm absorbers to more extreme ultra-fast outflows, hereafter UFOs) thought to originate from the accretion disk, the inner broad line region and/or the inner torus via a yet unknown physical process \citep{reeves_2003_outflowpds456, pounds_2003_pg1211, blustin_2005_aa, tombesi_2010_sampleufo, kaastra_2014, mehdipour_2015, nardini_2015, cappi_2016, parker_2018_b}.

To progress on both these topics is of extreme importance with wide ranging implications. For example, black hole spin encodes information about its growth history, may play an active role in setting relativistic jets and energetic outflows shaping the evolution of their host galaxy, determines the radiative accretion efficiency and sets the magnitude of some of the most extreme general relativistic phenomena observable in the Universe, such as gravitational redshift, light bending \citep{reynolds_2019_nat}. 
Similarly massive outflows at very high velocity (up to $\sim 0.3~\rm c$) which mere existence has been a challenge to standard theoretical models of wind formation may well be responsible for the so-called AGN feedback and explain the AGN-host galaxy co-evolution (e.g. \citet{tombesi_2019_wpusds,laha_2019_wpusds}, and references therein).

Remarkably, both phenomena (relativistic reflection and fast, massive outflows) are often seen together \citep{gallo_2019,walton_2019}. Consequently, understanding and disentangling their precise contribution has often been a matter of debate \citep{boller_2002, tanaka_2004, fabian_2009_nature, miller_2010, zoghbi_2011}. This is because the emission and absorption features, imprinted by partial covering, multiple ionization absorbers can mimic, within available data, those expected from reflection components, and vice versa \citep{done_2007, miller_2009_mnras,gallo_and_fabian_2011, gallo_and_fabian_2013,miller_2013_apj}.

In order to shed light on these studies, key is to obtain a combination of high resolution spectra (to detect lines) and high throughput (to fill in all energy channels), as the one provided by the X-IFU instrument onboard Athena \citep{barret_2018_spie}. It is therefore not a surprise that the two above topics are core science for Athena \citep{nandra_2013_sp,dovciak_2013,cappi_2013_sp}. X-IFU combines a better than 2.5 eV spectral resolution up to 7 keV, and a peak effective area of $\sim 1$ m$^2$ at 1 keV, and capabilities to observe with $\sim 100$\% throughput the brightest known AGN  \citep{barret_2018_spie}. 

Here we investigate for the first time the accuracy reached by X-IFU observations in measuring black hole spins, the geometry of the reflection and the parameters of the absorbers, considering realistic, though complex multi-component AGN spectra. In the next section, we first present our methodology to model, simulate and fit spectra (\S \ref{methodology}), present the results of a set of representative simulations highlighting some key parameters of the model (\S\ref{simulations}),  describe the prospects of measuring ultra fast outflows in high redshift objects (\S\ref{ufo}), introduce a method to estimate systematic errors due to calibration errors (\S\ref{calib}). The results are discussed in \S\ref{discussion}, where a comparison with similar studies is presented.

\section{Methodology}
\label{methodology}
\subsection{Model settings}
To simulate AGN X-IFU spectra, we assume an underlying continuum X-ray emission which consists of a hard cutoff power law component and its relativistically smeared ionized reflection, assuming a lamp post geometry for the irradiating source (\relxilllp\footnote{The model can be downloaded from \url{http://www.sternwarte.uni-erlangen.de/~dauser/research/relxill/}} in XSPEC) \citep{dauser_2013_mnras,garcia_2014_apj}. The free parameters of the model are the photon index ($\gamma$) of the incident continuum, the height of the lamp post (h, in units of gravitational radius, \rg$= GM/c^2$), the black hole spin parameter (a), and the inclination (Incl), ionization (\logxi), and iron abundance (\afe\ in solar units) of the accretion disk. The high-energy cutoff of the power law is  fixed to 300 keV, meaning that we do not have any curvature in the X-IFU energy range. When simulating the spectrum, the reflection fraction ({\it reflfrac}, hereafter \rf) can be either forced to a specified value (setting the model parameter {\it fixReflFrac} to 0) or computed self-consistently within the model and fixed to the lamp post value (setting {\it fixReflFrac} to 1), as defined in \citet{dauser_2016_aa}. The inner disk extends from the radius of the (spin dependent) innermost stable circular orbit up to 400 gravitational radii. The height of the compact source is constrained to lie within 3 and 10 \rg, consistent with the observations \citep{fabian_2009_nature,demarco_2013_mnras,emmanopulous_2014_mnras,gallo_2015_mnras}. Here we consider only positive black hole spin values, limited to 0.998.  

The primary power law plus disk reflected emission are then seen through three absorbers of varying column density (N$_{\rm H~1,2,3}$), ionization parameter ($\log \xi_{1,2,3} $) and covering factor ($cvf_{1,2,3}$). The absorber thought to be closer to the black hole has a higher column density and higher ionization, and the other two have non overlapping, though continuous, \nh\ and ionization parameters. Here we arbitrarily constrain the covering factor to range between 0.4 and 0.9. The absorber is modeled with \zxipcf\ in XSPEC. \zxipcf\ uses a grid of XSTAR photoionised absorption models (calculated assuming a micro-turbulent velocity of 200 km/s) for the absorption \citep{kallman_2001,reeves_2008}. Having the turbulent velocity a free parameter of the model will soon be implemented (C. Done, private communication). 

Reflection on cold distant neutral material is also accounted for and is also subject to obscuration by the two most distant, less ionized absorbers. Cold reflection is modeled using the {\it xillver} model \citep{garcia_2013_apj} (but see e.g. \citet{tanimoto_2019_apj} for a recent discussion on torus based models). Only the reflected component is computed setting {\it reflfrac}=-1. The power law index of the irradiating source is tight to the one of the primary emission (with again the same high energy cutoff set to 300 keV). The iron abundance of the reflector is set to 1 and its ionization parameter is set to 1 (\logxi=0), meaning an almost neutral reflector. It is known that the resolution of the {\it xillver} model, currently 17 eV at 6 keV \citep{garcia_2010_apj}, is significantly worse than the 2.5 eV X-IFU spectral resolution. However, for the prime focus of this paper, this is unlikely to be an issue, as we are interested primarily in the relativistically smeared reflection component, with broadening exceeding the resolution of the model. Reflection models with finer resolution, to fully exploit the X-IFU capabilities, may become available (J. Garcia, private communication).

\begin{table}
\caption{Model name, parameter and range of values (between Min and Max) assumed in the simulations (see text for details), where N$_{\rm{H}}$, $\xi$, height (h), and inclination (Incl) are given in units of 10$^{22}$cm$^{-2}$, erg cm$^{-1}$s, \rg, and degrees, respectively.}             
\label{table:1}      
\centering                          
\begin{tabular}{c c c c }        
\hline\hline                 
Model & Parameter & Min & Max  \\    
\hline                        
TBabs  & \nh\ & 0.01 & 0.05 \\      
\hline        
  & N$_{\rm{H 1}}$ & 0.3 & 0.6 \\      
$\rm zxipcf_1$ & $\log \xi_1$ & 0.5 & 1.5 \\ 
 &  $cvf_1$ & 0.4 & 0.9 \\ 
 \hline   
 &  N$_{\rm{H 2}}$ & 0.6 & 1.0  \\      
$\rm zxipcf_2$ &  $\log \xi_2$ & 1.5 & 3.0  \\ 
 &  $cvf_2$ & 0.4 & 0.9    \\ 
 \hline   
 &  N$_{\rm{H 3}}$ & 5 & 10 \\      
$\rm zxipcf_3$  & $\log \xi_3$ & 2.5 & 4.0 \\ 
  & $cvf_3$ & 0.4 & 0.9  \\ 
\hline                                   
& a & 0.05 & 0.95 \\
&$h$& 3.0 & 10.0 \\
 & $\gamma_{relxilllp}$ & 1.7 & 2.2 \\
$\rm relxilllp $& $\log \xi $ & 2.0 &\ldots \\
& \afe & 1.0 & 2.0 \\
& Incl & 30 & \ldots \\
& \rf & \ldots & 2.0 \\
\hline             
& \afe & 1.0 & \ldots  \\
$\rm xillver $& $\log \xi $ & 0 &\ldots \\
& $\gamma$ & $\gamma_{relxilllp}$ & \ldots \\
& Incl & 30. & \ldots \\
\hline       
\end{tabular}
\end{table}

\begin{figure*}
   \centering
   \includegraphics[width=0.49\hsize]{eeufspec.eps} 
   \includegraphics[width=0.49\hsize]{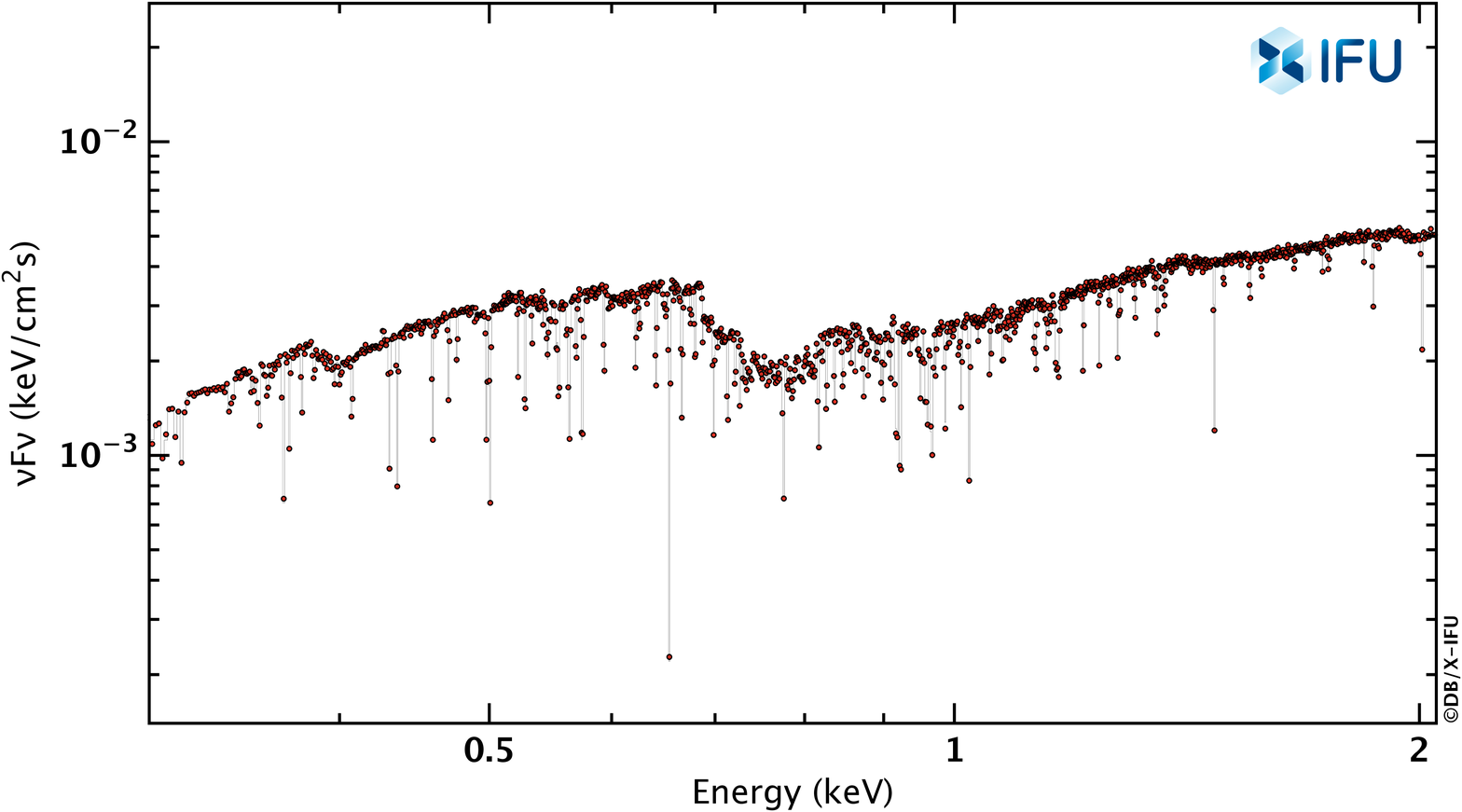} 
   \includegraphics[width=0.49\hsize]{powerlaw_wo_ref.eps}
    \includegraphics[width=0.49\hsize]{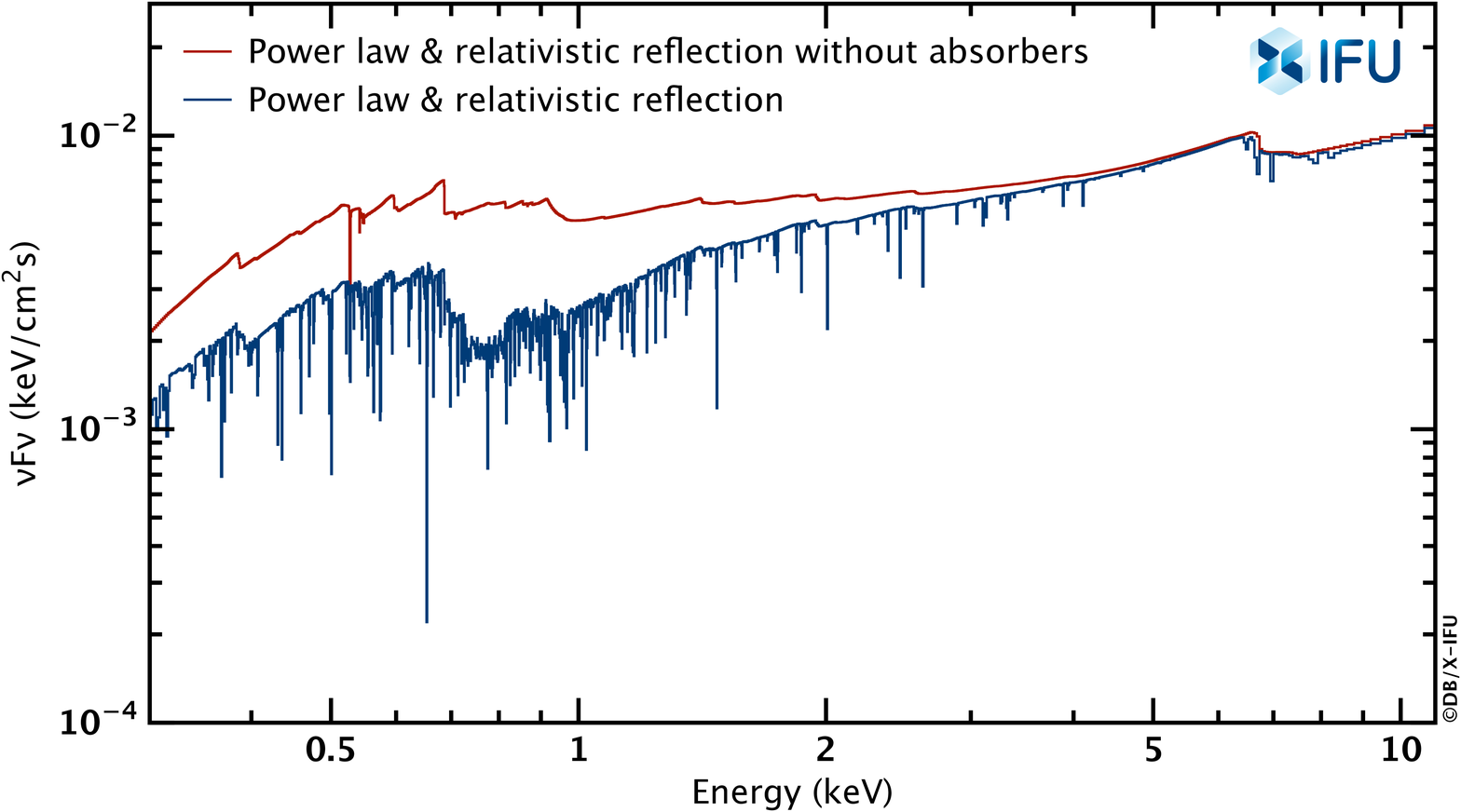} 
    \includegraphics[width=0.49\hsize]{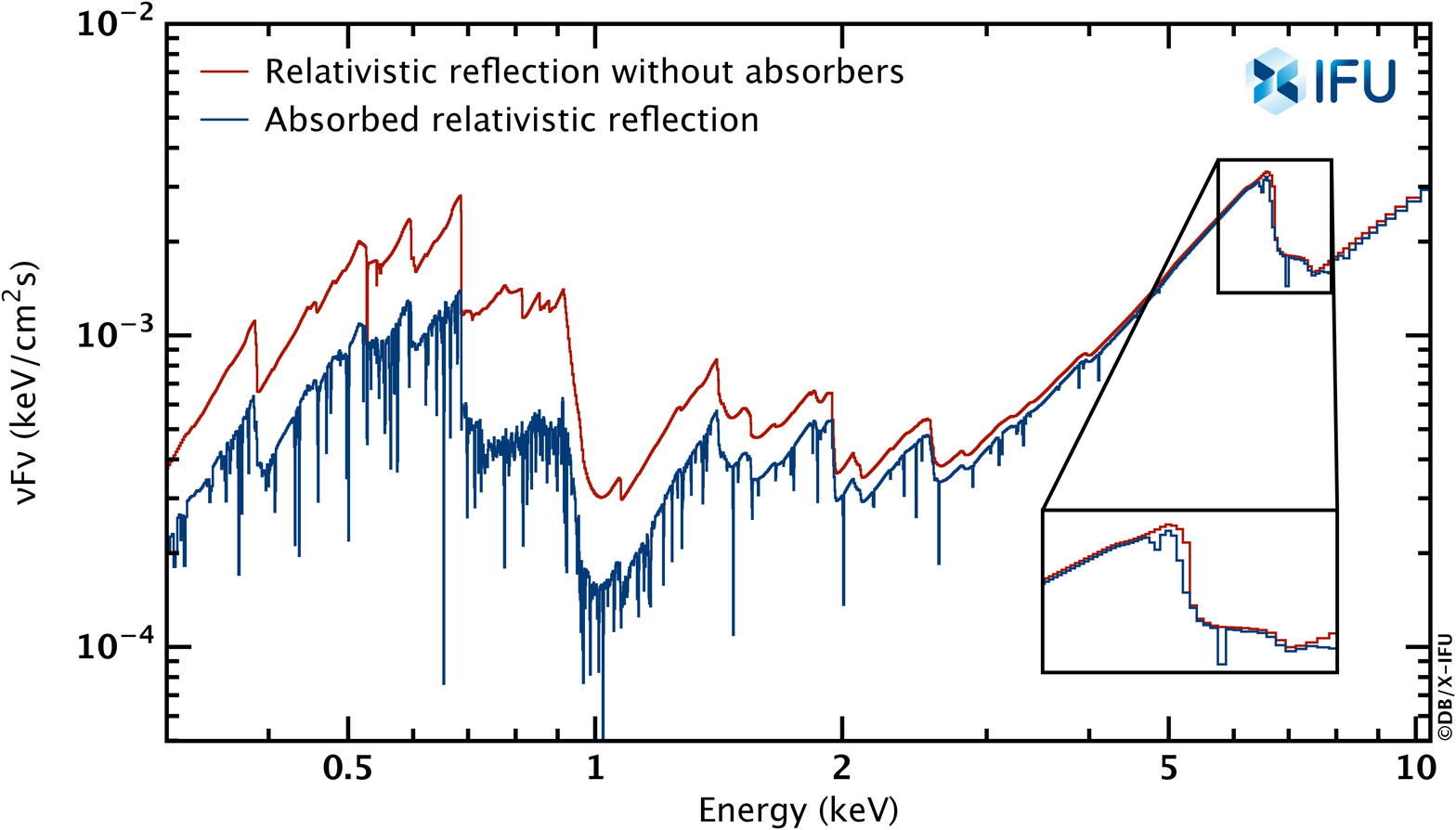} \includegraphics[width=0.49\hsize]{cold_ref.eps}
   
    \caption{Top-left: Simulated X-IFU spectrum for a black hole spin parameter of 0.65 for a \rf=2, \afe=2 and \logxi=2 with an inset around the FeK energy band. The source flux corresponds to a 1 mCrab source, and the integration time is set to 100 ks. Top-right: A zoom of the spectrum below 2 keV. The imprint of the absorbers on the various broad band emission components is shown in the subsequent panels: middle-left: the power law, middle-right: the power law plus the relativistic reflection component, bottom-left: the relativistic reflection component and  bottom-right: the cold reflection component from distant material. The different components are shown with and without the absorbers. The bottom lower left panel shows multiple bumpy features below 2 keV due to ionized reflection. These are key in constraining the black hole spin.}
         \label{eeufspec}
   \end{figure*}

We first consider a system inclination of 30 degrees and a redshift of all the components set to 0 (unless mentioned otherwise). Galactic absorption is modeled through TBabs \citep{verner_1993,wilms_2000}, with N$_{\rm H}$ allowed to vary between 1 and $5 \times 10^{20}$ cm$^{-2}$.  In XSPEC terminology, the model considered is $\rm TBabs \times (\rm zxipcf_1 \times zxipcf_2 \times zxipcf_3 \times relxilllp + zxipcf_1\times zxipcf_2\times xillver)$. The main parameters of the model are presented in Table \ref{table:1} together with their allowed range of variations. These values are estimated from \cite{delacalle_2010, tombesi_2010_sampleufo, laha_2014}. They can be considered representative of typical values of  nearby Seyfert 1 galaxies. An example of simulated X-IFU spectrum, highlighting the imprint of the various absorbers, and the contribution of the different reflectors is shown in Figure \ref{eeufspec}. Beside the forest of absorption lines present in the spectrum, it is worth noticing that the shape of the ionized reflection component shows multiple bumpy features below $\sim 2.5$ keV (see bottom left panel of Figure \ref{eeufspec}), which is key in constraining the black hole spin, in support to the constraints provided by the broad iron line above 6 keV. 

\subsection{Simulation setup and fitting}
The simulations are performed with the PyXspec interface to the XSPEC spectral-fitting program \citep{arnaud}. We use here a build in version of XSPEC  $12.10.1$. For simulations intended to sample the spin parameter space, we consider 50 values ranging from 0 to 0.995 in regular spacing. For all the other free physical model parameters, they are drawn from a uniform distribution bounded by their allowed interval of variations (listed in Table \ref{table:1}). The overall model is first normalized to correspond to an absorbed flux equivalent to a 1 mCrab source in the 2-10 keV range (or an absorbed flux of $2 \times 10^{-11}$ \ergs). The sample size of 50 is commensurate with the number of known unabsorbed (typically type 1) AGN of similar brightness, which currently populate the Athena mock observing plan\footnote{The Athena mock observing plan can be downloaded from \url{http://www.isdc.unige.ch/athena/document-repository/category/192-general-interest.html}}, and as can be found in common catalogues such as the 3XMM-DR8 \citep{rosen_2019}, Chandra \citep{evans_2019}, Swift \citep{oh_2018} catalogues. 

The normalization of the cold reflection {\it xillver} component is realistically assumed to be one fifth of the \relxilllp\ component. As a sanity check of this assumption, we fitted the 2-10 keV spectrum with a simple power-law model plus three Gaussian lines, and the best fit equivalent width of the emission lines at $\sim$ 6.2, 6.4 and 7.1 keV were 70, 120 and 20 eV, respectively. These values are broadly consistent with typical values measured for the FeK line redshifted, neutral and K$_{\beta}$ components, e.g. \cite{guainazzi_2006_an,nandra_2007_mnras, delacalle_2010}. The spectra are then generated using the latest response matrices of the X-IFU \citep{barret_2018_spie} and the latest background files\footnote{Available for download from \url{http://x-ifu.irap.omp.eu/resources-for-users-and-x-ifu-consortium-members/}. The response files used here are named XIFU\_CC\_BASELINECONF\_2018\_10\_10, which correspond to the configuration of the X-IFU presented at the Instrument Preliminary Requirement Review.}. Note that the background rate for a point source with an extraction radius of 5 arc seconds is less than $2 \times 10^{-4}$ counts/s and is negligible in the simulations (one mCrab source generates about 100 counts/s). For grouping the spectral bins, we consider the optimal binning scheme of \citet{kaastra_2016} using the ftools {\it ftgrouppha}. The scheme accounts in particular for the energy dependent spectral resolution of the instrument and the statistic of the spectrum (narrower bins near high count regions and wider bins near low count regions).  Depending on the model parameter simulated (e.g. slope of the power law, absorbers column density), the 1 mCrab count rate varies from $\sim 50$ counts/s to $\sim 120$ counts/s over the considered X-IFU fitting energy range (0.3 keV to 11.5 keV). When grouped, the mean energy bin width is less than 3 eV. There are 16 free parameters for the model considered here, and more than 6000 degrees of freedom (for a source of 1 mCrab brightness). 

We use the so-called \cstat\ metric in fitting the spectra \citep{cash_1979_apj,kaastra_2017_aa}, (see however section \ref{appendix_a} for an illustrative example of biases introduced by the use of $\chi^2$ statistic). To fit a spectrum, as with real data we will not know what the model parameters will be, we do not initialize the fit with the input model parameters. Instead, we draw tens (up to 50) sets of randomly distributed parameters in their allowed interval of variations. For the normalization of the primary emission and cold reflection components, we draw two numbers from a uniform distribution bounded as $\pm 50$\% of the input normalizations. To speed up the fitting, we constrain the fit to converge in 50 iterations, with a critical change in the fit statistic $\Delta \rm cstat=0.1$, assuming that any better fit will be found during the error computation (on all free model parameters). As these starting parameters may be far off from the input spectral parameters, the fit may not reach an acceptable solution before its 50 iterations and is simply ignored (of the 50 initial sets of parameters, at least one set leads to an  acceptable fit to launch the error computation). This method has the advantage that it sweeps well over the parameter space, as to avoid the fit to get trapped into a local minimum.  

For computing the errors on the best fit parameters, we consider the set of best fit parameters which provided the lowest \cstat. For the parameter of interest, the positive and negative errors are computed by varying its value around its best fit value, freezing it, and fitting the spectrum with all the other free parameters allowed to vary. The value is incremented until it exceeds a critical threshold ($\Delta \rm{cstat}=2.706$ for 90\% confidence level errors). The parameter value at $\Delta \rm{cstat}=2.706$ is obtained through interpolation of the $\Delta  \rm{cstat}$ curve. This is equivalent to the recommended {\it steppar} procedure in XSPEC. If a new best fit is found along the error computation, the procedure aborts and then restarts on the first free parameter from the newly found best fit. Computing the errors further sweeps over the parameter space, and is often used to get away from local minima in the fitting statistics, e.g. \cite{hurkett_2008}. With the method to initialize the fit described above, considering the simulation 1 described below, the mean decrease of \cstat\ over 50 simulations along the error computation is $\sim 0.1$ (for a mean value of $\sim 6405$), indicating that the global minimum was likely found, hence the best fit. This is further supported by computing the goodness of the fit. Following \cite{kaastra_2017_aa}, we compute the goodness of the fit from the expected value and expected variance of the \cstat. Again, in simulation 1 described hereafter, all measured \cstat\ values for the best fits are all within $\pm 2\sigma$ of their expected values, indicating that the spectral model is acceptable. A visual inspection of the $\Delta \rm{cstat}$ was applied to check on the behavior of some sensitive model parameters, such as the black hole spin.  

\section{X-IFU spectral simulations in representative configurations}
\label{simulations}
In the simulations, we assume an effective integration time of 100 ks (unless mentioned otherwise), with the implicit assumption that all model parameters remain constant within that duration (e.g. the height of the compact irradiating source, the parameters of the absorbers, etc.).

\subsection{The most conservative case: \rf=1, \afe=1 for a 1 mCrab source (configuration 1)}
To demonstrate the power of the X-IFU to constrain black hole spins, the very first simulation to be conducted assumes the most conservative case, in which the iron abundance and the reflection fraction are both set to 1, and we consider a mildly ionized disk with the ionization parameter set to $\log \xi$=2. We assume a 1 mCrab source. We simulate 50 spectra with positive spins ranging from 0 to 0.995 in regular spacing, to make sure that the whole spin range is covered. The mean error on the best fit parameters for the three absorbers and the reflection component are listed in Table \ref{table:2}. As can be seen, the statistical error on the spin parameter is on average $\le 0.1$, and the error on the height of the irradiating source is $\sim 0.3$\rg.

\subsection{Another conservative case: fixed lamp post geometry with \afe=1 for a 1 mCrab source (configuration 2)}
\begin{figure*}
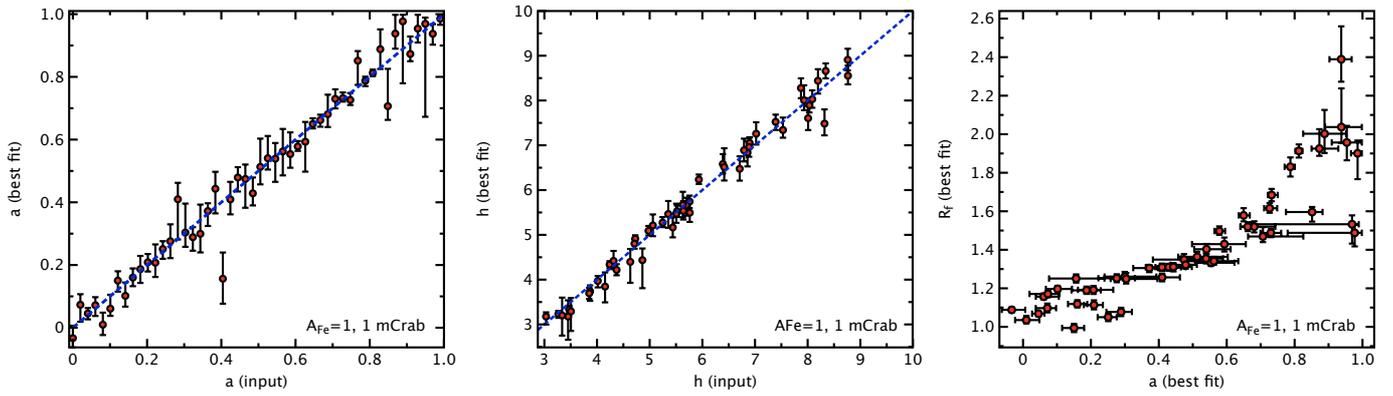

   \centering
       \includegraphics[width=0.33\hsize]{a.eps} \includegraphics[width=0.33\hsize]{h.eps}\includegraphics[width=0.33\hsize]{fracref_vesus_a.eps}
      \caption{The best fit parameters versus the input parameters for the case of a 1 mCrab source observed for a 100 ks X-IFU observation. The parameters of the reflection component are computed by the model to the predicted value of the geometrical configuration in the lamp post geometry \citep{dauser_2016_aa} and then left as a free parameter of the fit.  50 spins are generated to sample the range 0 to 0.995, while the height of the X-ray source is drawn at each spin from a uniform distribution bounded between 3 and 10 \rg. The iron abundance \afe\ is set to 1. This the so-called configuration 2 simulations. Left: the spin parameter, middle: the height of the compact source and right: the reflection fraction versus the fitted spin parameters. The errors are computed at the 90\% confidence level for variation of one single parameter. As predicted by the model, \rf\ tends to increase with the black hole spin: the scattering at high spins being due to the height of the X-ray source allowed to vary between simulations.}
         \label{conf2}
   \end{figure*}

In this other simulation run, the parameters of the reflection component are computed by the model to the predicted value of the current parameter configuration in the lamp post geometry \citep{dauser_2016_aa}, namely the height of the irradiating source and the spin. \rf\ is then left as a free parameter of the fit. The iron abundance is conservatively set to 1 and the ionization parameter of the accretion disk remains set to 2. We again consider 50 spectra with positive spins ranging from 0 to 0.995 in regular spacing, while for each spin, $h$ is drawn from a uniform distribution between 3 and 10 \rg. The mean error on the best fit parameters for the three absorbers and the reflection component are listed in Table \ref{table:2}. The best fit parameters (a, $h$, \rf) are shown in Figure \ref{conf2}. The reflection fraction computed from the model is everywhere smaller than 2, but larger than in the most conservative case discussed above. It goes from $\sim 1.2$ for low spin values up to $\sim 1.9$ at the highest spins. The higher reflection fraction at high spins compensates for the increased smearing of the reflection features, likely explaining why the error bars on the spin remain similar across the spin range. As listed in Table \ref{table:2} the accuracy of the fitted spin values has a mean error of $\sim 0.05$ across the spin range considered, while the height of the irradiating source is accurate to $\sim 0.2$\rg.

\subsection{Setting \rf=2, \afe=2 for a 1 mCrab source (configuration 3)}
Although its physical origin has often been debated (but see the hypothesis on radiative levitation by \cite{reynolds_2012}, iron over abundance and large \rf\ have often been reported from AGN X-ray spectra, with values reaching 10 and 5, respectively, in the most extreme cases \citep[and references therein]{fabian_2009_nature, risaliti_2013, parker_2018}. Similarly, iron over abundance is also inferred from fitting binary black hole spectra \citep{garcia_2018_aspc}, with values several times solar being routinely found. Here we assume a reflection fraction of 2 and iron over abundance by a factor of 2 at maximum, which may not be considered such an extreme case after all. \rf\ is fixed in faking the spectra, and then left as a free parameter of the fit. 50 sets of the remaining 15 parameters are drawn from their uniform distribution, with the spin ranging from 0 to 0.995 in regular spacing. The mean error on the best fit parameters for the three absorbers and the reflection component are listed in Table \ref{table:2}. The best fit parameters against the input values of the model are presented in Figure \ref{conf3}. There are several noticeable features in this figure. First the spin parameter is very well constrained, and as expected, as the smearing increases with the spin, the error on the spin recovered increases towards the highest spin values (this is not compensated by a larger \rf\ as it is fixed to 2 in the simulations). Nevertheless, the mean error on the spin parameter is $\lesssim 0.05$ across the range of spins considered. Second, in the lamp post geometry, the height of the compact source is also very well constrained with a mean error of the order of $\sim 0.2 $ \rg. \rf\ as recovered by the fit is shown in Figure \ref{conf3_ref}. \rf\ is again determined with an accuracy of $\sim 2$\%. Similarly, the power law index of the hard irradiating source has a negligible error (0.003), showing no bias against its input value (see section \ref{appendix_a} for the bias that using $\chi^2$ as the fitting metric would introduce). Finally, the parameters of the three absorber components are very well recovered, most notably the ionization parameter (typical errors of $\sim 2$\% over their range of variation). As expected at the highest ionization, the best fit error increases as there are less lines in the spectra to hook the fit. A degeneracy between the \nh\ and the covering factor may be expected (a higher \nh\ and a smaller covering factor can be found as equal to a smaller \nh\ and a higher covering factor). Despite this, the recovered values are all consistent or very close to their input values, demonstrating the power of high resolution and high throughput spectroscopy to decipher multiple components, narrow and broad, in complex AGN spectra. Not shown on this summary plot, it should be added that the normalization of the two reflection components and the galactic \nh\ are also consistent with their input values. 

\begin{table*}[!t]
\caption{Mean 90\% confidence level one sided errors from the best fit spectral parameters of the various configurations of the simulations. Only the parameters of the three absorbers and the ones of the reflection component are listed. The mean one sided errors are the mean of the positive and negative errors. Note that the positive error on the spin is bounded by the maximum spin value of 0.998. Configuration 1b corresponds to the same simulations as for configuration 1 but using the WFI responses \citep{meidinger_2018_spie}. Configuration 3b is identical to configuration 3 but a 5\% energy independent systematics has been added to the data. The integration time for configuration 1 to 4, and 7 to 9 is 100 ks. For configuration 5 and 6, it is 25 ks. The source intensity is 1 mCrab in configuration 1 to 4 and 0.5 mCrab in configuration 5 to 9. Configurations 7 to 9 are identical to configuration 3 but the iron abundance, the system inclination and the disk ionization are allowed to vary, as indicated in Table \ref{table:3}.}             
\label{table:2}      
\centering                          
\begin{tabular}{cccccccccccccc}       
\hline\hline                 
Conf. & $\Delta$N$_{\rm{H 1}}$ & $\Delta\log\xi_1$ & $\Delta cvf_1$ & 
$\Delta$N$_{\rm{H 2}}$ & $\Delta\log\xi_2$ & $\Delta cvf_2$ & 
$\Delta$N$_{\rm{H 3}}$ & $\Delta\log\xi_3$ & $\Delta cvf_3$ &
$\Delta h$ & $\Delta a$ & $\Delta \gamma$ & $\Delta$ \rf \\   
\hline                        
1 & 0.011 & 0.013 & 0.010 & 0.052 & 0.036 & 0.032 & 0.498 & 0.024 & 0.027 & 0.296 & 0.083 & 0.003 & 0.027 \\
1b & 0.044 & 0.053 & 0.049 & 0.164 & 0.074 & 0.081 & 1.579 & 0.063 & 0.097 & 0.592 & 0.148 & 0.004 & 0.042 \\
2 & 0.012 & 0.014 & 0.012 & 0.056 & 0.036 & 0.034 & 0.475 & 0.022 & 0.023 & 0.217 & 0.048 & 0.003 & 0.036  \\
3 & 0.013 & 0.018 & 0.015 & 0.058 & 0.036 & 0.034 & 0.595 & 0.027 & 0.026 & 0.178 & 0.049 & 0.003 & 0.038 \\
3b & 0.016 & 0.022 & 0.019 & 0.065 & 0.035 & 0.030 & 0.521 & 0.021 & 0.023 & 0.180 & 0.045 & 0.003 & 0.042 \\
4 & 0.043 & 0.050 & 0.040 & 0.174 & 0.086 & 0.077 & 1.574 & 0.066 & 0.068 & 0.595 & 0.168 & 0.009 & 0.119 \\
5 & 0.043 & 0.049 & 0.036 & 0.169 & 0.076 & 0.068 & 1.506 & 0.066 & 0.078 & 0.610 & \ldots & 0.009 & 0.066 \\
6 & 0.041 & 0.044 & 0.035 & 0.112 & 0.061 & 0.055 & 2.162 & 0.077 & 0.082 & 0.604 & \ldots & 0.008 & 0.063 \\
\hline
7 & 0.027 & 0.028 & 0.026 & 0.090 & 0.045 & 0.040 & 0.753 & 0.031 & 0.031 & 0.495 & 0.113 & 0.004 & 0.070 \\
8 & 0.027 & 0.027 & 0.030 & 0.074 & 0.045 & 0.046 & 0.767 & 0.044 & 0.041 & 0.543 & 0.107 & 0.004 & 0.051 \\
9 & 0.021 & 0.025 & 0.017 & 0.063 & 0.037 & 0.028 & 0.660 & 0.026 & 0.029 & 0.370 & 0.118 & 0.005 & 0.054 \\
\hline       
\end{tabular}
\end{table*}

\subsection{Setting \rf=2, \afe=2 for a 0.1 mCrab source (configuration 4)}
Given the small uncertainties on the spin determination in Figure \ref{conf3}, it is tempting to investigate how the X-IFU would perform on sources ten times weaker, opening the possibility to explore the spin distribution of weaker seyfert galaxies and/or more distant quasars. So we repeated the simulations above, but assuming a source corresponding to a flux of 0.1 mCrab, equivalent to $2 \times 10^{-12}$ \ergs\ (2-10 keV), allowing the integration time to increase to 150 ks (to compensate partly for the ten time lower brightness). The mean error on the best fit parameters for the three absorbers and the reflection component are listed in Table \ref{table:2}. The results of the simulations are shown in Figure \ref{conf4} for a sample of 20 simulations. As can be seen, the spin determination is less accurate, yet the error bars on the spin parameter are typically $\sim 0.17$, with the same tendency as above for lower spin values to be determined more accurately, as expected given the sharper line features then.

\subsection{Recovering the height of the compact source on shorter timescales (configuration 5)}
Assuming that the spin of the black hole is known, it is worth investigating how the height of the compact source could be measured on timescales of the order of 25 ks, thought to be commensurable to the characteristic variability timescale of these sources, i.e. typically bright nearby Seyfert galaxies \citep{ponti_2012}. We consider here a 0.5 mCrab source with a spin parameter of 0.5. We assume a conservative iron abundance for the disk, an ionization $\log\xi =$ 2 and \rf\ computed self-consistently from the \relxill\ model in the lamp post geometry, allowing the height of the irradiating source to vary between 3 and 10 \rg. We simulate 25 X-IFU spectra, with different corona heights, power law index, absorber parameters. We let the reflection fraction a free parameter of the fit. The mean error on the best fit parameters for the three absorbers and the reflection component are listed in Table \ref{table:2}. The best fit results for the compact source height and the reflection fraction are shown in Figure \ref{conf5}. On 25 ks timescale, the accuracy with which $h$ can be measured is about $0.6$ \rg, while \rf\ is determined with an accuracy of 5\%. Combining such spectral information, with timing analysis (e.g. measuring time lags), would enable a detailed mapping of the accretion geometry around the black hole.

\subsection{Recovering the parameters of the UFOs (configuration 6)}

We have shown previously the capability of X-IFU to separate the three absorbers imprinting on a complex reflection spectrum. Next we focus on the third high density, high ionization absorber, when blue shifted. In low resolution AGN X-ray spectra, these absorbers manifest as narrow Fe K-shell blue shifted absorption lines from Fe XXV/XXVI, with inferred radial velocities between 0.03 - 0.3 c \citep{tombesi_2010_sampleufo, gofford_2013}. As the blue shift increases, the strongest high energy absorption lines due to iron get shifted towards higher energies and separate clearly from the relativistic iron line, but fall in an energy range where the effective area of the X-IFU decreases sharply. Yet, as shown below, constraints on the parameters of this absorber are expected to come also from the low-energy absorption lines, which are also well resolved by the X-IFU. 

We carry out a set of simulations, considering a fixed black hole spin (0.5), the reflection computed in a fixed lamp post configuration, blue shifting the third absorber component with velocities ranging between $-0.3$ and $-0.05$ the speed of light, with the parameters of the three absorbers again drawn from uniform distributions bounded in their interval of variations listed in Table 1 (the redshift of the first two other absorbers remains 0). We keep the same \zxipcf\ model for the absorber, but we smear it with a gaussian (gsmooth in XSPEC), to account for an additional broadening of 1000 km/s at 6 keV, consistent with UFO observations \citep{tombesi_2011}. This will  smear out the absorption features, thus reducing the benefit of a high resolution spectrometer for that component, while it remains crucial to separate the other absorbers. The XSPEC model used is $\rm TBabs \times (\rm zxipcf_1 \times zxipcf_2 \times (zxipcf_3 \otimes gsmooth) \times relxilllp + zxipcf_1\times zxipcf_2\times xillver)$.

We assume a 0.5 mCrab source and 25 ks for the spectrum integration time, again because one would be interested to probe those UFOs on the shortest possible timescales. The mean error on the best fit parameters for the three absorbers and the reflection component are listed in Table \ref{table:2}The results of the fit for the case of a turbulent velocity of 1000 km/s are shown in Figure \ref{conf6}. As can be seen, the accuracy by which the absorber parameters are recovered has decreased, due to the smearing, although the redshift (i.e. velocity) of the absorber is measured with a very high accuracy. 

\subsection{Varying other key parameters: iron abundance, system inclination, disk ionization (configurations 7 to 9)}
\label{otherparameters}
The above simulations have all considered fixed iron abundance (\afe=1 or 2), fixed inclination (30 degrees) and fixed ionization (\logxi=2) of the disk. It is interesting to see how those parameters can be constrained, if varying and left as free parameters of the model, and which impact this would have on the spin measurement accuracy. We have simulated 3 sets of 25 spectra for a 0.5 mCrab source, the spin covering 0 to 0.995 and an exposure time of 100 ks (\rf=2, \afe=2). The range of allowed variations for \afe, system inclination and disk ionization are listed in Table \ref{table:3}. The mean error on these specific parameters is given as the last column of \ref{table:3}. The mean error on the other best fit parameters are reported in Table \ref{table:2} in the lines corresponding to configurations 7 to 9. As can be seen, those parameters are recovered with high accuracy, most notably the varying disk ionization parameter which is one of the key parameter defining the reflection spectrum. Leaving those parameters, as free parameters of the fit does degrade the accuracy by which the spin and the height of the irradiating source are measured by about a factor of 2.
\begin{table}
\caption{Allowed range of variations for the the iron abundance, system inclination and the ionization parameter of the disk. The mean one sided 90\% error on these parameters are listed from the fitting of 25 simulated spectra. Each simulation considers a 0.5 mCrab source observed for 100 ks and a reflection fraction of 2, and a uniform distribution of the varying parameter in their allowed range of variations. The errors on the other parameters are listed in table \ref{table:2}, including errors on the spin and the height of the irradiating source.}             
\label{table:3}      
\centering                          
\begin{tabular}{c l c c }        
\hline\hline                 
Configuration & Parameter & Range & Mean error \\    
\hline                        
7 & \afe & 1-10 & 0.269 \\           
8 & Inclination (deg.) & 10-70 &  0.123 \\     
9 & $\log\xi$ & 1-3 & 0.015 \\           
\hline\hline     
\end{tabular}
\end{table}

\subsection{Soft excess}
The measurement of the reflection parameters relies on the broad band coverage of the X-IFU, not only the iron K$\alpha$ line (6-7 keV), but also the relativistically smeared features below 2 keV or so (see Figure \ref{eeufspec}, bottom left panel). As a matter of fact, a simple test of ignoring the data below 2 keV in the fit shows that it would significantly reduce the accuracy on the best fit parameters for the complex model considered here. Below 2 keV is an energy range in which a soft excess (in addition to the underlying power law component) is often requested by the data, with a significant contribution to the total flux. Its origin is still debated. Hypothesis include an extra warm Comptonization, complex partial covering, disk blackbody emission, reprocessed reflection component and even a relativistically blurred high-density reflection \citep{magdziarz_1998, crummy_2006,gierlinski_2006, mehdipour_2015, petrucci_2018, middei_2019,garcia_2019_apj}.  Let alone the fact that X-IFU with its unprecedented sensitivity at $\sim 1$ keV will provide critical insights on the origin of the soft excess, for this paper, it is  important to test whether the presence of such a soft component, if not related to relativistic reflection could affect the accuracy by which the reflection parameters and the black hole spin are measured. 

We have thus simulated a set of 50 spectra in configuration 3 (\rf=2, \afe=2, \logxi=2), adding to the input model a steep power law with a photon index ranging between 2.5 and 3.5. Alternative models for the soft excess would be a thermal comptonization, {\it comptt}-like or a blackbody model. The exact shape of the soft excess does not matter here: what matters being the number of counts added on top of the reflection spectrum. The corresponding XSPEC model is then $\rm TBabs \times (\rm zxipcf_1 \times zxipcf_2 \times zxipcf_3 \times (relxilllp + powerlaw) + zxipcf_1\times zxipcf_2\times xillver)$. The power law normalization is such that the 0.5-2 keV flux of the power law component is conservatively set to 50\% of the total flux in that energy range. This can be considered conservative as on average it leads to a 0.5-2 keV unabsorbed flux higher than the 2-10 keV flux (by $\sim 10$\%), while they are generally found to be comparable, e.g. \citet{miniutti_2009_mnras}. As we are only interested in the errors on the spin and height, the fit starts with the input model parameters and the errors are computed on these two parameters only. We find a mean error on the spin of $\sim 0.13$ and $\sim 0.36$ \rg\ on the height. This is to be compared with the values of $\sim 0.05$ and $\sim 0.18$ reported in table \ref{table:2}. As expected, the accuracy on the spin and height measurement has decreased, because the broad features of the relativistic reflection below 2 keV are diluted by the soft excess. They remain however acceptable, in the conservative setting, used for the model.

\section{Beyond the local Universe: a z=2.5 AGN}
\label{ufo}
We have demonstrated above that even for moderately bright sources (0.1 mCrab)  X-IFU is able to characterize both the absorption components and the reflection spectra simultaneously and with great precision. We now want to investigate how well it could perform on even more distant (i.e. fainter) sources, considering that redshifting significantly the spectrum would bring the absorption/emission features closer to the peak of the effective area of the X-IFU (see Figure \ref{fig_aeff}), thus compensating partly the reduction of flux. 

The model we consider is a simplification of the model above in which the first two absorbers are merged into one. The XSPEC model becomes $\rm TBabs \times (\rm zxipcf_1 \times (zxipcf_2 \otimes gsmooth) \times relxilllp + zxipcf_1\times xillver)$. The covering factor of the two absorbers is set to an intermediate value of 0.75. We thus first consider an AGN with a flux of $2 \times 10^{-13}$ \ergs\  (i.e. 0.01 mCrab) \citep{georgakakis_2013,martocchia_2017_aa,dadina_2018,baronchelli_2018_mnras}. The normalization of the cold reflection component relative to the relativistic remains one fifth. The reflection component is computed in a fixed lamp post configuration assuming a black hole spin of 0.5. {Note that no meaningful constraints can be derived on the spin at this flux level}. We further assume a height of the irradiating source of 4 \rg\ with \afe=1 and \logxi=2.  The redshift of the first absorber is set to the redshift of the source, while a blue shift (between -0.3 and -0.05) is added to the second absorber. A Gaussian velocity broadening of 1000 km/s normalized at 6 keV (rest frame, the index of the gaussian smoothing function in XSPEC is assumed to be 1) is assumed for the high density absorber. A simulated spectrum corresponding to an exposure time of 100 ks is shown in Figure \ref{fig_distant_agn}, to highlight the imprints of the two absorbers on the spectrum, leaving a forest of absorption lines that will be crucial to measure the redshifts.

To be more quantitative, we simulate 10 spectra with the redshift of the source allowed to vary between 2.4 and 2.6, and a large velocity broadening of 3000 km/s for the high density absorbers. The fit is performed between 0.3 and 3.5 keV.  Both the source redshift and the redshift of the absorber are then left free in the fit. In the framework of this simplistic model, unsurprisingly the redshift of the source would be determined with a high accuracy due to the prominent redshifted iron line produced by the distant reflector (statistical error less than $\sim 0.001$). Despite the larger velocity broadening, the blueshift of the high density absorber would be measured with a statistical error much less than $\sim 0.01$. The best fit X-ray redshifts for the source and the UFO are plotted in Figure \ref{fig_redshifts}. More detailed simulations are warranted with added complexity to the model, but as discussed by \cite{martocchia_2017_aa}, who pushed the flux limit by yet another order of magnitude, snapshot X-IFU observations may reveal the presence of outflows, imprinting absorption lines around the peak of the effective area of the X-IFU. Such observations may as well provide the X-ray redshift of the source and would probe the occurrence rate of outflows, their temporal variability and their link with the kpc-scale outflows running through the interstellar medium, right at the golden epoch of AGN-galaxy evolution at redshift above 2 \citep{martocchia_2017_aa}.  

\section{Accounting for calibration uncertainties}
\label{calib}
It can be predicted that a high resolution spectrometer such as the X-IFU will be challenging to calibrate. In this section, we investigate how uncertainties in the instrument effective area may affect the present results, which are provided so far just with statistical errors, not accounting for any systematic errors. The first test we performed was by allocating a 5\% systematic error to the spectra for the configuration 3 (\afe=2, \rf=2). The mean error on the best fit parameters are reported in Table \ref{table:2} for the so-called configuration 3b. It is encouraging to see that the impact on the accuracy of the best-fit parameters is very small. Next we introduce on a more detailed analysis accounting for the X-IFU calibration requirements.

Per its performance requirements, the X-IFU is required to have a knowledge on the shape of the effective area curve better than 3\% ($1\sigma$), across the 0.2 to 10 keV range. In total, with the mirror assembly, the requirement is not to exceed 5\% ($1\sigma$, on axis). In addition, for the X-IFU the normalisation of the effective area should be known with an absolute error lower than 4\% at 1 keV (still at $1\sigma$), with a contribution from the mirror of less than 6\%. In a first step, we first restrict our exercise to X-IFU, ignoring the additional uncertainties arising from the mirror. 

At low energies, the X-IFU quantum efficiency is determined by the transmission of the optical/thermal blocking filters and their supporting meshes \citep{barbera_2018_spie, barret_2018_spie}. Those filters are made of Polymide, Aluminum and Aluminum Oxyde. On the other hand, at high energies the quantum efficiency derives from the transition edge sensor absorber thickness, currently 1.7 $\rm{\mu m}$ of gold and 4.2 $\rm\mu m$ of Bismuth, e.g. \cite{peille_2018_spie}. Here we follow the Monte Carlo simulation approach introduced by \cite{drake_2006_spie} and followed recently by \cite{cucchetti_2018_spie}. We first generate a large number of auxiliary response files (1000) whose shape remains within the envelope of the $\pm 3$\% maximum allowed shape deviation. We do this by bounding the thicknesses of the filters and the absorbers, e.g. the thickness of the gold absorber is drawn from a conservative unbounded normal distribution of $\sigma=0.13$ $\rm \mu m$ centered around 1.7 $\rm \mu m$ (\cite{drake_2006_spie} assumed the distribution to be truncated at $\pm 1 \sigma$). The energies at which the maximum deviation is computed are 0.5 keV and 10 keV instead of 0.2 and 12 keV. We ignore uncertainties around the edge of the response. Once the overall shape of the effective area curve is determined, its normalization is drawn from an unbounded normal distribution of mean 1, and $\sigma=0.04$ (the same normalization applies not only at 1 keV but throughout the whole energy band, see Figure \ref{fig_aeff}).

There are two possible approaches to estimate the errors linked to the uncertainties in the instrument response. \citet{drake_2006_spie} proposed to start from a single simulated spectrum generated from the nominal response file, and fit it with the newly generated response files. With this approach, the statistics being the same, it better highlights the perturbations induced by the calibration uncertainties. On the hand, the statistics of the one single spectrum to fit may have an important role in the results, as we deal with spectra with millions of  counts (i.e. the fit may converge to the same best fit if the changes in the response shape are not significant enough). At the opposite, \citet{cucchetti_2018_spie} faked one spectrum per newly generated response files and fit each of them with a single, nominal response file. In both cases, we wish to compare the distribution of the best fit parameters with the distribution expected from pure Poisson statistics. We estimate the latter by faking the same number of spectra with the nominal response file, fitting them and recording the best fit parameters (i.e. the usual way of estimating best fit errors from Monte Carlo simulations).

As in real life, the end user of the X-IFU will likely be provided with one single response file to fit the data, affected by calibration uncertainties, here we prefer the approach of \cite{cucchetti_2018_spie}. For the sake of this exercise, we simulate spectra for the so-called configuration 3 (1 mCrab, 100 ks, \afe=2, \rf=2), and an intermediate spin value of 0.5. It is important to note that before all fits, the simulated spectra are binned optimally, accounting for the response file used \citep{kaastra_2016}. All fits are performed between 0.3 and 11.5 keV, and as we are interested in assessing only the distribution of best fit parameters, they all start with the model input values.

The distribution of best fit parameters for the main reflection parameters are compared in Figure \ref{fig_res_cal_uncertainties} (the spin, the height of the irradiating source, the reflection fraction and the power law index). If systematic calibration errors were to be important, the two distributions should differ significantly, with the distribution from Poisson statistics being narrower than the one accounting for both Poisson statistics and calibration uncertainties. As can be seen from Figure \ref{fig_res_cal_uncertainties}, for that particular case, the calibration errors considered here (at the X-IFU level only) are small, in particular for the spin which is of prime interest here (the mean error on the spin increases from $\sim 0.02$ to $\sim 0.03$). It is also interesting to note that the calibration uncertainties, as modeled here, are not introducing any biases in the distributions. 

If we were considering that all calibration errors at the mirror assembly plus instrument level would arise from X-IFU alone (i.e. changing 3\% to 5\% and 4\% to 10\%), we have repeated the  simulations above, regenerating another set of 1000 response files. The systematic errors increase as expected but remain small for the spin parameter (increasing from $\sim 0.02$ to $\sim 0.04$). The error on the height goes at the same time from $\sim 0.2$ to $\sim 0.4$ \rg. 

It should be anticipated that the calibration errors may affect differently different parameters, depending on whether they are sensitive to the low energy part of the response or to the high energy part, or if they relate to a continuum component or relate to a component with discrete features, such as absorption/emission lines. A more detailed analysis of the calibration requirements for X-IFU, and their impact on Athena driving science cases (such as the detection of the missing baryons in the warm hot intergalactic medium) will be devoted to a follow-up paper, in which we will also compare our method of assessing the systematics with the method of \citet{drake_2006_spie}. 

\section{Discussion}
\label{discussion}
We have demonstrated for the first time in a quantitative way the power of high resolution spectroscopy to decipher complex multi-component AGN X-ray spectra. Next, we briefly discuss on the advances permitted by X-IFU on probing black hole spins, accretion-ejection physics, the issue of over iron abundances found in AGN X-ray spectra and conclude by a comparison with similar feasibility studies, using different instrumental settings. 

\subsection{New insights on the black hole spin distribution}

Measuring the distribution of spins of a large (greater than $\sim$ 50) sample of super massive black holes may tell us about their growth channels, including the relative contributions of mergers versus prolonged accretion \citep{berti_volonteri_2008,dovciak_2013}. Models predict that mergers would lead to a flat spin distribution while prolonged disk-mode accretion would end up with black holes spinning rapidly. Of the few tens of AGN with known spin values, the distribution is peaked towards high spins \citep{vasudevan_2016_mnras,reynolds_2019_nat}. This has been argued as a consequence of the fact that, in a flux limited sample, black holes with higher spins, accreting at the same rate are likely to be over represented because of their higher radiation efficiency ($\eta = 0.057$ for a non-spinning black hole to $0.32$ for maximal spin a = 0.998). Another bias that may be present in the current data comes from the fact that higher black hole spins lead to larger \rf\ \citep{dauser_2014_mnras}. 

{In the near future, the eROSITA All-Sky Survey, reaching a 2-10 keV sensitivity limit about two orders of magnitude lower than the previous HEAO-1 All-Sky Survey \citep{piccinotti_1982_apj} will increase the number of objets at z larger than 1 and brighter than 0.1 mCrab from the handful known today to several hundreds \citep{comparat_2019_mnras}. Interestingly enough, recent very deep (4 Ms) Chandra exposures on the Chandra Deep Field South indicate that high z and nearby objects may share similar spectral properties, in particular by the presence of a broad iron line \citep{baronchelli_2018_mnras}. Thus, with the accuracy reached on the spin measurement at the 0.1 mCrab flux level (0.17 in the configuration 4 above), and thanks to its unbiased sensitivity to measure low spins (even for relatively high source heights of 10 \rg), the results presented in this paper demonstrate that the X-IFU carries the potential to provide unprecedented constraints on the intrinsic black hole spin distribution, up to z$\sim$1-2. Such a result would have wide-ranging implications, such as for constraining the black hole growth models \citep{berti_volonteri_2008}, but also for correcting luminosity functions or  constraining black hole population synthesis models as discussed by \cite{vasudevan_2016_mnras}.}

\subsection{New studies of accretion and ejection flows}
The X-IFU will open a spectroscopic window to address strong gravity accretion physics and probe outflows over a range of physical parameters for the corona/reflection component of accretion disks and of AGN-driven winds, down to unprecedented short time scales and faint source fluxes. This will enable new, currently mostly unpredictable, type of studies of accretion and ejection phenomena. We address qualitatively a couple of such cases, considering that more extensive simulations would be required. 

X-IFU will be able to probe the height of the X-ray compact source with respect to the accretion disk down to less than a fraction of \rg\ and on time-scales comparable to the X-ray source variability time scales (see Figure \ref{conf5}). In a way comparable to what is currently done in coronal mass ejections from the sun, e.g. \citet{Gou_2019}, it is possible that the X-ray coronae in AGN are also formed by magnetic reconnection events on top of an accretion disk. This will lead to strong flaring, massive coronal loops and particle acceleration. The spectral information provided by X-IFU, combined with reverberation lags between the direct and reflected emissions will probe the geometry and corona-disk structure down to the innermost regions of the accretion disk, where most of the energy is released \citep{dovciak_2013,wilkins_2016_simlags, zoghbi_2019_wpusds}.

Measuring with great precision the different parameters of AGN and QSO-driven winds, such as their ionization parameter, column density and velocity, is key to understand whether such winds have a sufficiently high mechanical power (typically 0.5 per cent of the bolometric luminosity) to provide a significant contribution to AGN feedback \citep{hopkins_2010_mnras,fabian_2012_nature,cappi_2013_sp}. Kinetic energies being proportional to the v$^3$, precise measurements such as the ones shown in Figure \ref{conf6}, i.e. yielding typical errors less than few \%, are mandatory. But beside this classic argument, another new opportunity, introduced in \cite{cappi_2013_sp} is the possibility offered by X-IFU to measure not only their line shifts (i.e. velocity) but their line profile with unprecedented precision, again down to either short time scales and faint sources, i.e. in nearby Seyferts and more distant QSOs. Such information would be key to constrain the launching site and mechanisms for the winds (see \citet{dorodnitsyn_2009_mnras} for detailed simulations of such profiles, and \citet{chartas_2016_apj} for a tentative application to Chandra data). 

 \cite{done_2007} and \cite{nardini_2015} have shown that the broad FeK emission lines combined with the strong absorption features at higher energies seen in some bright nearby AGN may well be interpreted as P-Cygni line profiles produced by a spherically symmetric wind or shell. Similar P-Cygni profiles should be seen for all absorption lines but with different shapes and different time variations for the different absorbers. In addition, a realistic flow will be rather radially extended with a distribution of kinematical, ionization, and dynamical properties along the line of sight, leading to even more complex absorption profiles \citep{proga_and_kurosawa_2010, giustini_and_proga_2012}. Simulating such complex spectra with X-IFU goes beyond the scope of this paper, but clearly X-IFU holds the potential to provide key insights into the winds properties.

\subsection{Iron over abundance, disk inclination and ionization}
Inferred iron over abundances from X-ray reflection spectroscopy is one of the most intriguing results, which casts some doubts on the reported spin values given the tight relation between reflection parameters and iron abundances. This has motivated the revision of reflection models towards densities above the currently used values: those densities being expected in the vicinity of black holes \citep{garcia_2016_mnras,garcia_2019_wpusds}. Application of high density models to a few selected objects has already shown that the iron abundance recovered was significantly lower than the one obtained with lower-density disk reflection models, e.g. \cite{tomsick_2018_apj,jiang_2019_mnras}. Those models which are currently under development have clear signatures at energies below 1 keV. In particular, the enhancement of free-free heating in the atmosphere of the disk, increasing with increasing density leads to a soft excess. These high density models will be easily testable with X-IFU, which will measure the iron abundance down to solar, together with the reflection component with high accuracy (see Tables \ref{table:2} \& \ref{table:3}). 

We have also shown (\S\ref{otherparameters}) that the disk inclination and ionization will be well constrained. This is very important because inclination measurements could allow comparison of inner disk inclinations to those for the host galaxy stellar disk, thereby put constraints on the way AGN are fueled. Material propagating inward through the galactic disk or via minor mergers are expected to leave imprints on their average respective alignment \citep{middleton_2016}. 

Understanding how much the reflection component shall be ionized is also an open and debated issue. The FeK line profile is, in principle, carrying  sensitive information on the disk ionization state, but in practice it is often degenerate with the other free parameters of the line profile. As a result, the soft energy band is key to constrain the amount of ionization for the reflection component as shown in the lower left panel of Figure \ref{eeufspec} where the disk soft emission becomes quickly very significant at intermediate up to high ionization levels. As a note of caution, it is worth stating that in our lamp-post model, the ionisation is assumed to be constant radially. Ideally it should be calculated self-consistently with the radial density to account for the centrally peaked illumination expected in a relativistic accretion disk model, see \cite{martocchia_2002, svoboda_2012} and in particular \cite{kammoun_2019} for a consideration of this effect, including also X-IFU simulations. 

\subsection{Comparison with other instruments and simulations}

Feasibility studies have so far been carried out, considering X-ray spectra with limited spectral resolution, $\sim 100$ eV at 6-7 keV, e.g. as provided by XMM-Newton EPIC instruments \citep{strueder_2001_aa}, combined with hard X-ray data enabling to sample the smooth Compton reflection bump above 10 keV, e.g. as provided by NuSTAR \citep{harrison_2013_apj}. We briefly discuss here how these studies compare to those presented in this paper.

\cite{kammoun_2018} have conducted a similar analysis to ours, simulating XMM-Newton EPIC-PN and NuSTAR spectra in the range of 1 to 3 mCrab fluxes. Beside relativistic reflection, they considered a warm absorber and two layers of partially covering neutral absorbers, cold reflection and thermal emission from the galaxy, thus introducing complexity in their spectral model similar to ours. The success rate of measuring their spin is about 50\% (over 60 fits). This rate increases to 100\% for spins larger than 0.8 and a lamp-post height lower than five gravitational radii (because this configuration imprints stronger, easier to detect, relativistic distortions to the spectrum, see also \cite{choudhury_2017}). On the other hand, the success rate goes to zero if the height of the irradiating source is at a distance larger than 5 \rg. As demonstrated above, X-IFU can measure spins all across the range investigated, and this even for small reflection fractions and iron abundance of 1, and source height up to 10 \rg. Interestingly, \cite{kammoun_2018}, considering two of their failed simulations, with the height of irradiating source at 11 and 18 \rg, noticed that Athena WFI simulations would not be more successful (despite the much improved statistics), concluding that this was likely due to the non proper sampling of the reflection hump above 10 keV. Because the Compton hump is not properly sampled by X-IFU either, it may be more likely due to a too low reflection fraction (due to the large source heights considered). We have repeated the simulations of Configuration 1 (\afe=1 and \rf=1) using the WFI response files\footnote{The response files were downloaded from \url{http://www.mpe.mpg.de/ATHENA-WFI/response_matrices.html} and the date of the version used is November 2017. See \cite{meidinger_2018_spie} for a recent description of the WFI instrument.} and found that the accuracy by which WFI recovers the reflection parameters is a factor of $\sim 2$ less than X-IFU, despite the higher effective area of the WFI at high energy ($\sim 25$\% around 6 keV). Taking advantage of its better spectral resolution, at the same time, the X-IFU recovers the parameters of the absorbers with error bars that are between a factor of 3 to 4 smaller (see Conf. 1b in Table \ref{table:2}). 

\cite{bonson_2016} have considered a model based on \relxill\ only (i.e. without absorbers and cold reflection, see \cite{choudhury_2017} for a discussion on their fitting scheme). They have simulated  spectra with XMM-Newton EPIC-PN and NuSTAR for the brightest seyferts, and found that the spin parameter could only be well measured for the most rapidly rotating super-massive black holes (i.e. $a > 0.8$ to about $\pm 0.10$). The error on the spin would reach $\sim 0.30$ at $a=0$ for \rf=5, a value not considered here. Interestingly enough in their simulations, they found that the addition of NuSTAR hard X-ray data did not improve the spin determination, see Figure 7 of \citet{bonson_2016}. At first sight, the simulations performed here do not seem heavily impacted by the lack of hard X-ray data (above 10 keV), possibly because there is sufficient information across the X-IFU band pass, in particular in the soft X-ray band where the ionized reflection component contributes significantly. 

Following up on this, it is worth noting that the power law index is extremely well constrained within our simulations, despite the model complexity. The assumption of a straight power law in the X-IFU band pass is correct for any plausible high-energy cutoffs (above tens of keV). \cite{garcia_2015_apj} showed that the high-energy cutoff up to even 1 MeV can be constrained using X-ray data below 100 keV by the sole modeling of the reflection component. This is due to the fact that the reflection spectrum, imprinted by fluorescent lines and other atomic features, depend sensitively on the shape of the emission spectrum of the irradiating source. We have repeated the simulations in configuration 3 (1 mCrab, 100 ks, \rf=2, \afe=2), leaving the energy cutoff as a free parameter, allowing it to vary between 50 and 200 keV (drawing the initial 50 values from a uniform distribution). Such a cutoff range is consistent with the latest Swift/XRT-NuSTAR observations of type 1 AGN \citep{molina_2019_mnras}, see also \citep{ricci_2017_apjs} and references therein. The mean 90\% confidence level error on the energy cutoff derived from the X-IFU simulations is $\sim 15$ keV over the 50 to 200 keV range, with a tendency for the errors to increase at the upper end of the range. This indeed suggests that meaningful constraints can be obtained on the high energy cutoff from the X-ray data alone. This also means that combining X-IFU data with comparably sensitive hard X-ray data, e.g. from the High-Energy X-ray Probe (HEX-P), proposed as a complementary mission to Athena \citep{madsen_2018_spie} would set very tight constraints on the reflection parameters, by measuring precisely both the energy cutoff and the Compton hump. It is also worth noting that the shape of the Compton hump, being independent on parameters such as the iron abundance or the disk ionization would help in removing model degeneracies, in case data are more complex than the ones simulated here (as they will likely be).  

Finally, \cite{choudhury_2017} have tested the \relxill\ model with simulated NuSTAR data, and assumptions more extreme than ours, e.g. \rf\ values up to 10, iron over abundance up to 10 also. They have also considered NuSTAR spectra accumulated over 100 ks and delivering between 1 and 10 millions counts. For the model considered here and a source of 1 mCrab, the rate expected in one NuSTAR module is $\sim 0.5$ counts/s between 3 and 70 keV\footnote{Response files were downloaded from \url{https://www.nustar.caltech.edu/page/response_files}}, meaning that the fluxes they considered would correspond to 20 to 200 mCrab for X-IFU: a flux regime not explored in this paper (and in which there are just a couple of AGN). They found that better constraints are obtained for smaller height of the irradiating source and larger reflection fractions, yet, the errors that they obtained in the most favorable conditions exceeds by at least one order of magnitude the one we obtain in our more realistic and complex setting. To take an example, for a spin parameter input of 0 and \rf=1 and $h=3$\rg, the 90\% dispersion among the simulations goes from $\sim -0.5$ to $\sim 0.25$, while for sources 200 times fainter the X-IFU would reach an error of $\le 0.1$. 

\subsection{Comparison with {\it XRISM}-Resolve}
{The X-ray Imaging and Spectroscopy Mission (XRISM), a JAXA/NASA collaborative mission, with ESA participation is expected to launch around 2021 \citep{tashiro_2018_spie}. It will carry Resolve, a soft X-ray spectrometer, which combines a lightweight soft X-ray telescope paired with a X-ray calorimeter spectrometer, to provide non-dispersive 5-7 eV energy resolution in the 0.3-12 keV bandpass. Opening the way to broad band high-resolution X-ray spectroscopy, which we have seen being critical for the science of interest in this paper, it is interesting to compare how Resolve will perform compared to X-IFU, despite its lower effective area (about a factor of $\sim 45$ at 1 keV and a factor of $\sim 5$ at 6 keV), and this, at least for the brightest objects. For the sake of this simple comparison and focussing on the spin measurements, we have simulated a 5 mCrab source in the so-called conservative configuration 1 above (\rf=1, \afe=1). We have generated 50 spectra with a constant spin spacing between 0 and 0.995 and with an integration time of 100 ks. Setting a favorable case, we have ignored the background and initiated the fit to the model input parameters. The error on the spin parameter has then been computed. In about $\sim 15$\% of the simulations, the fitted spin pegged at the hard limit. The mean error on the spin is $\sim 0.3$. With the same settings, the mean error on the spin from X-IFU observations would be $\sim 0.04$.}

{To summarize, the comparison with the three  feasibility studies similar to the one presented here, as well as the comparison with the XRISM-Resolve above, clearly demonstrate the advances the X-IFU will permit over existing and future instrumentations. }
\section{Conclusions}
The Athena X-IFU, as currently designed, is predicted to be transformational in many field of astrophysics, so will Athena overall, by the complementarity of its science payload \citep{nandra_2013_sp,barret_2013_sf2a,barcons_2017_an,guainazzi_2018_arxiv}. Here we have demonstrated the rather unique and outstanding capabilities of X-IFU for probing AGN spins, AGN surroundings, accretion disk physics, winds and outflows from local to more distant AGN, {using a state of the art reflection model in a lamp post geometrical configuration}. The leap in sensitivity provided by X-IFU derives from its excellent spectral resolution, high throughput and broad band coverage. More feasibility studies of this type, possibly combining spectral-timing analysis, extending the range of models to be tested, {the range of reflection geometries} and the range of objects to be considered (e.g. X-ray binaries) should be performed to further assess and quantify its unique capabilities. The methodology presented here may also serve this purpose.

\section{Appendix A: Biases in $\chi^2$ fitting}
\label{appendix_a} 
\begin{figure*}
   \centering
\includegraphics[width=0.33\hsize]{a.eps} \includegraphics[width=0.33\hsize]{h.eps}\includegraphics[width=0.33\hsize]{gamma.eps}
\includegraphics[width=0.33\hsize]{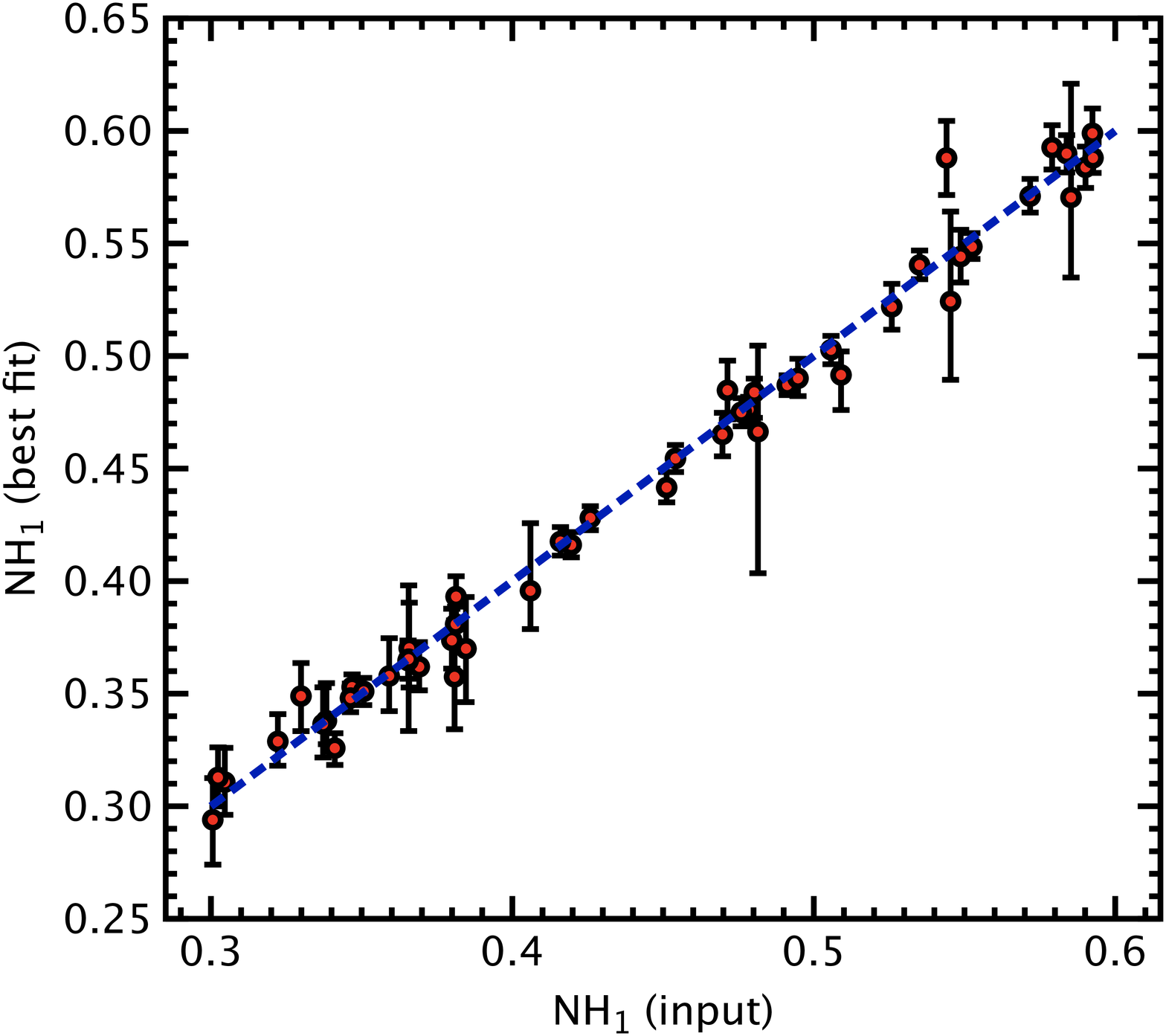} \includegraphics[width=0.33\hsize]{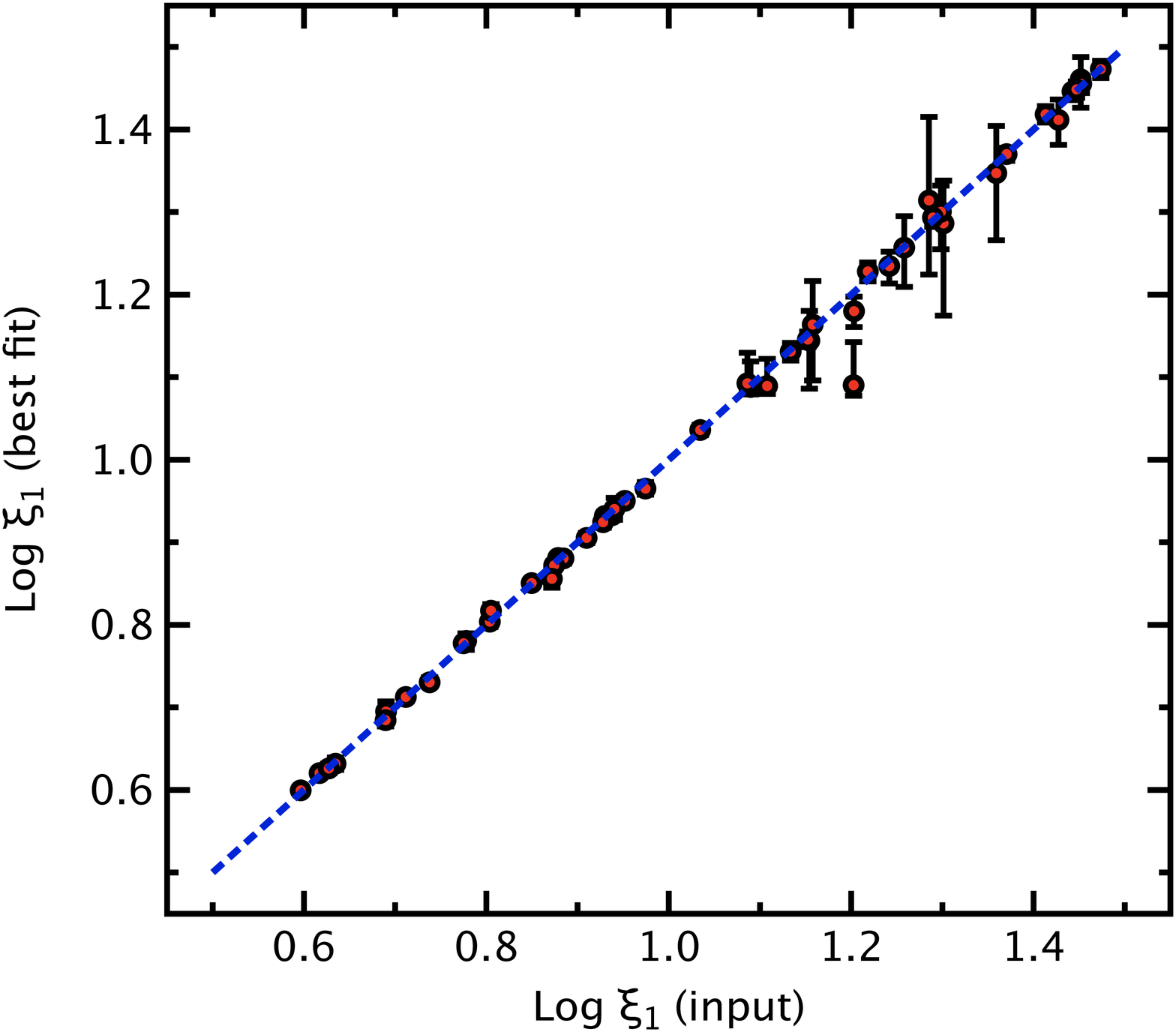}\includegraphics[width=0.33\hsize]{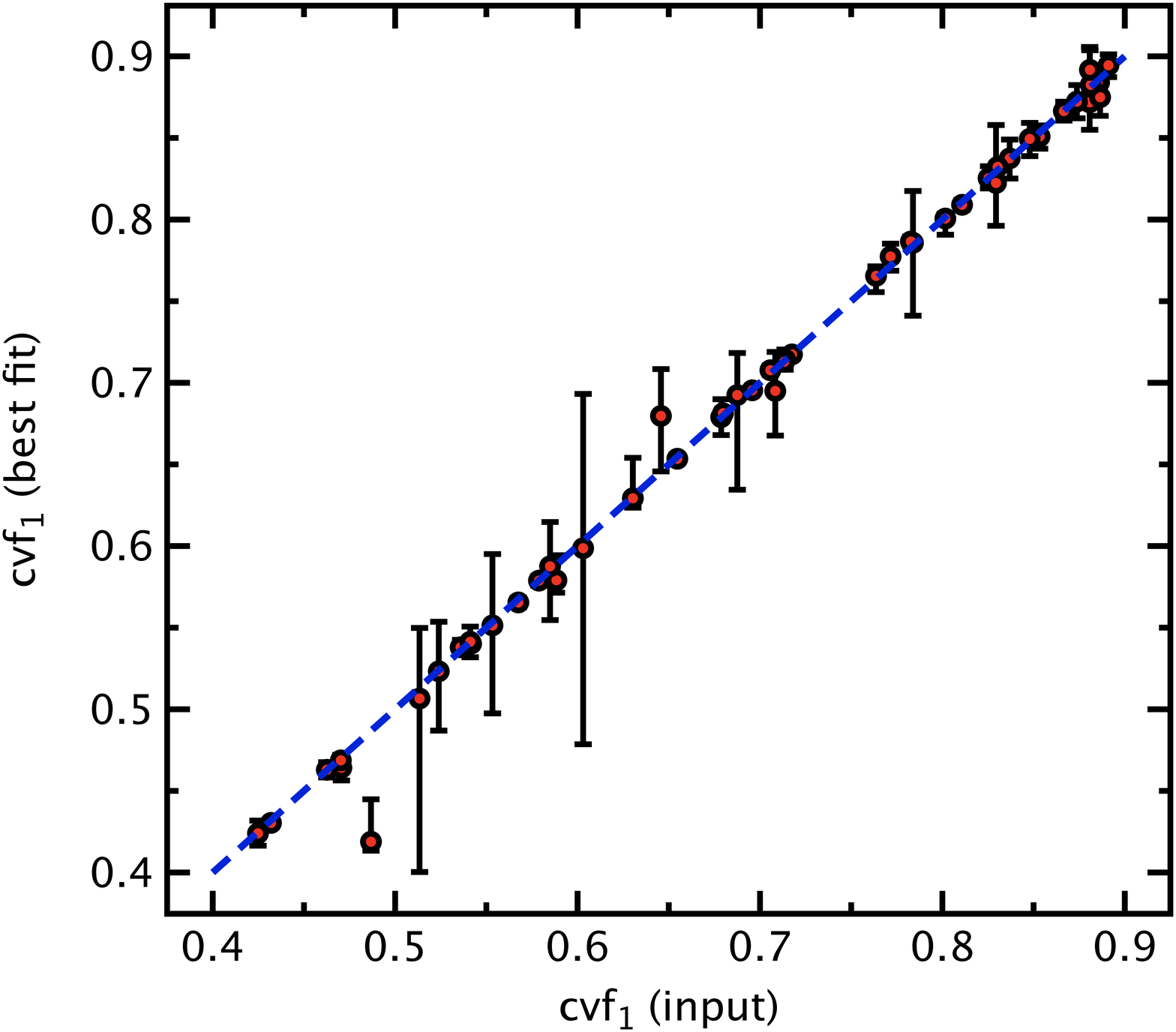}
 \includegraphics[width=0.33\hsize]{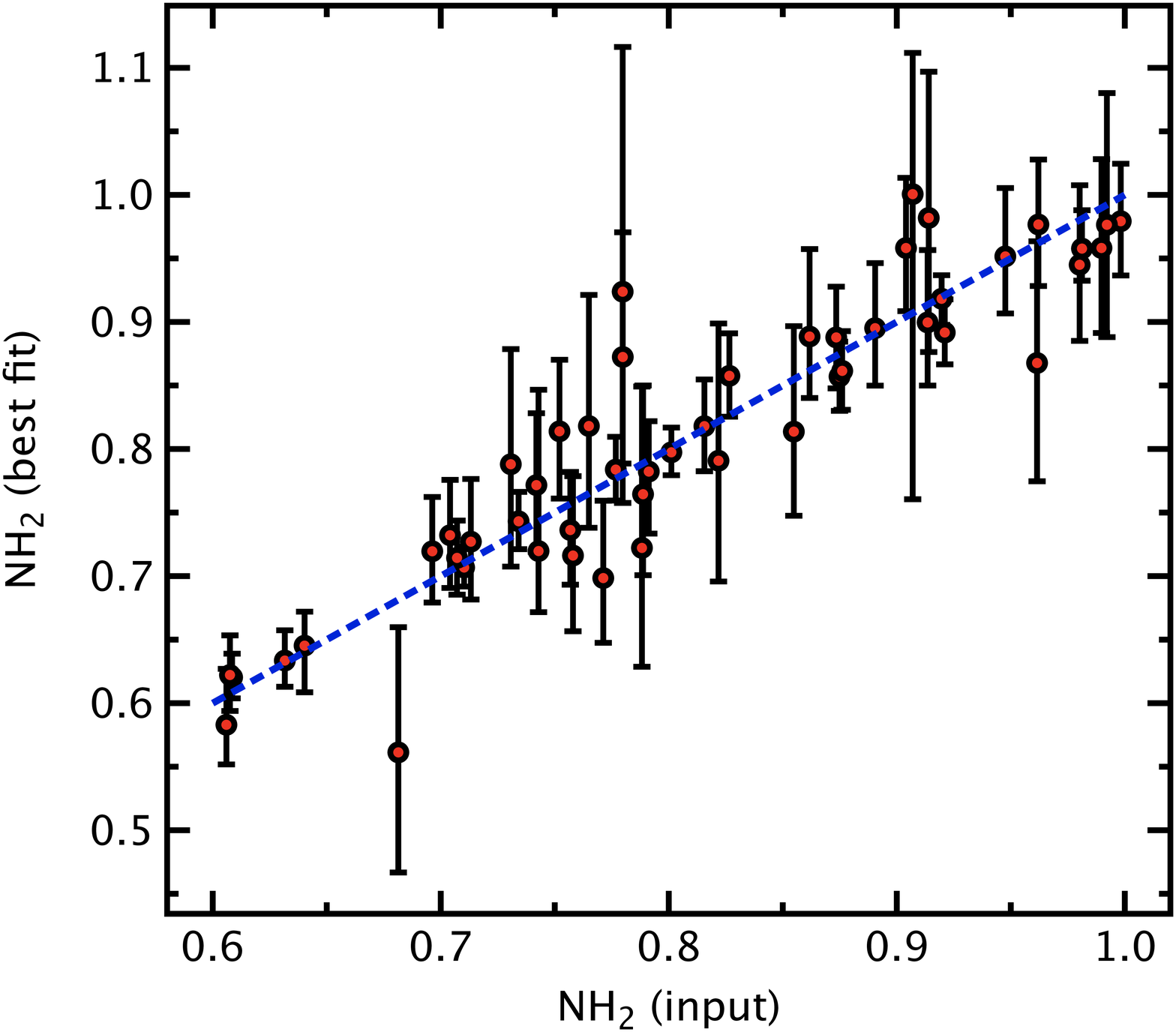} \includegraphics[width=0.33\hsize]{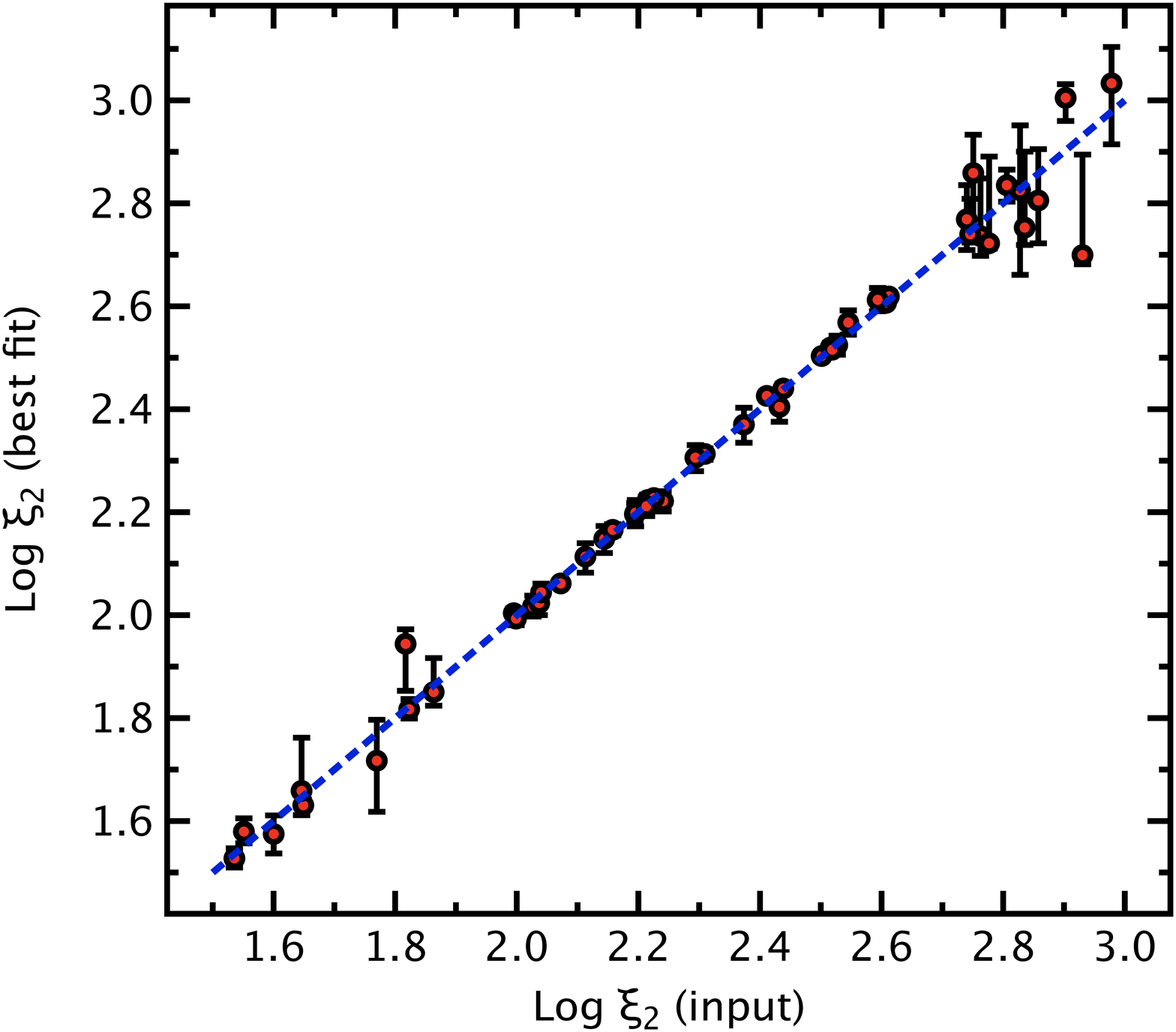}\includegraphics[width=0.33\hsize]{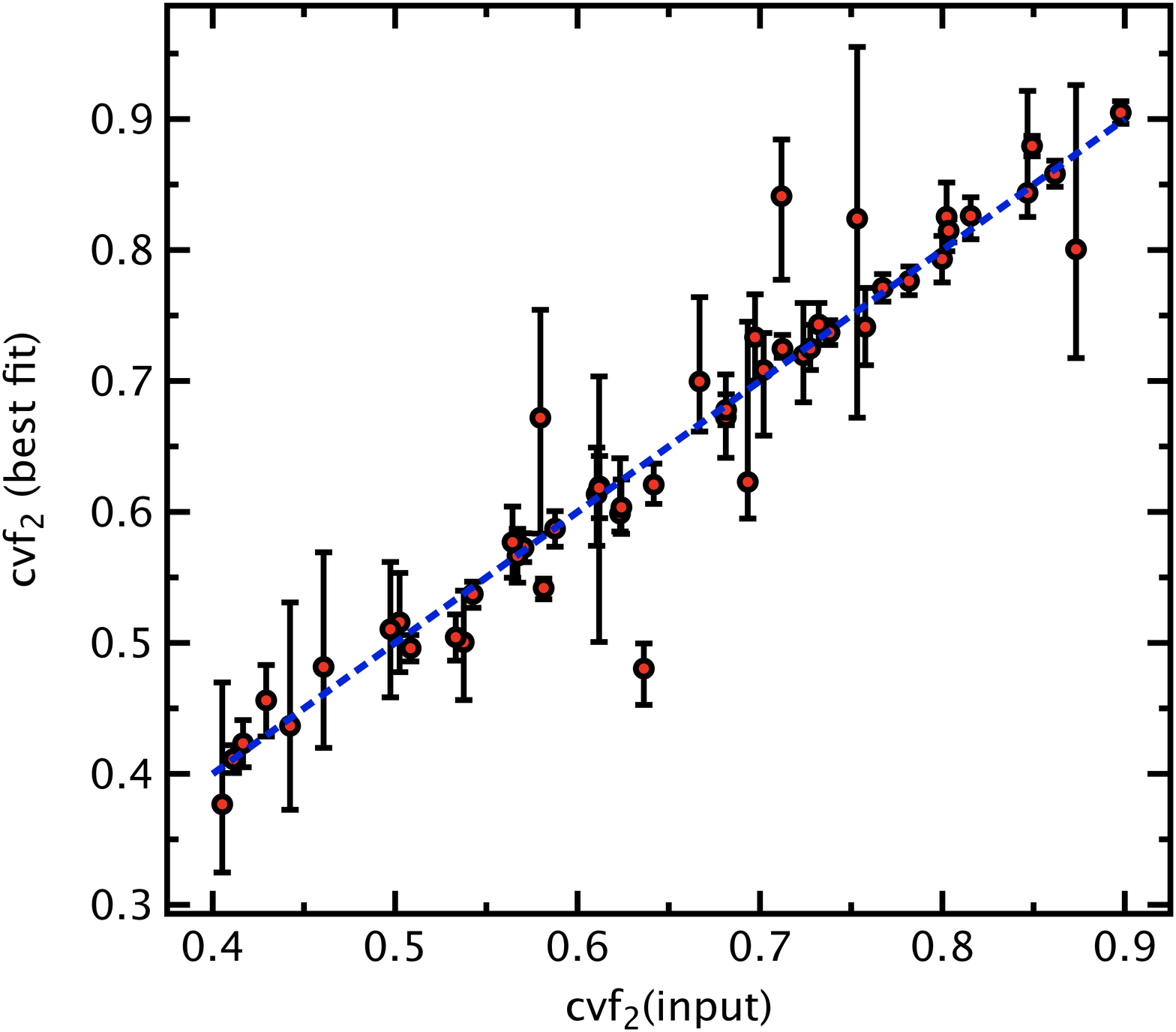}
 \includegraphics[width=0.33\hsize]{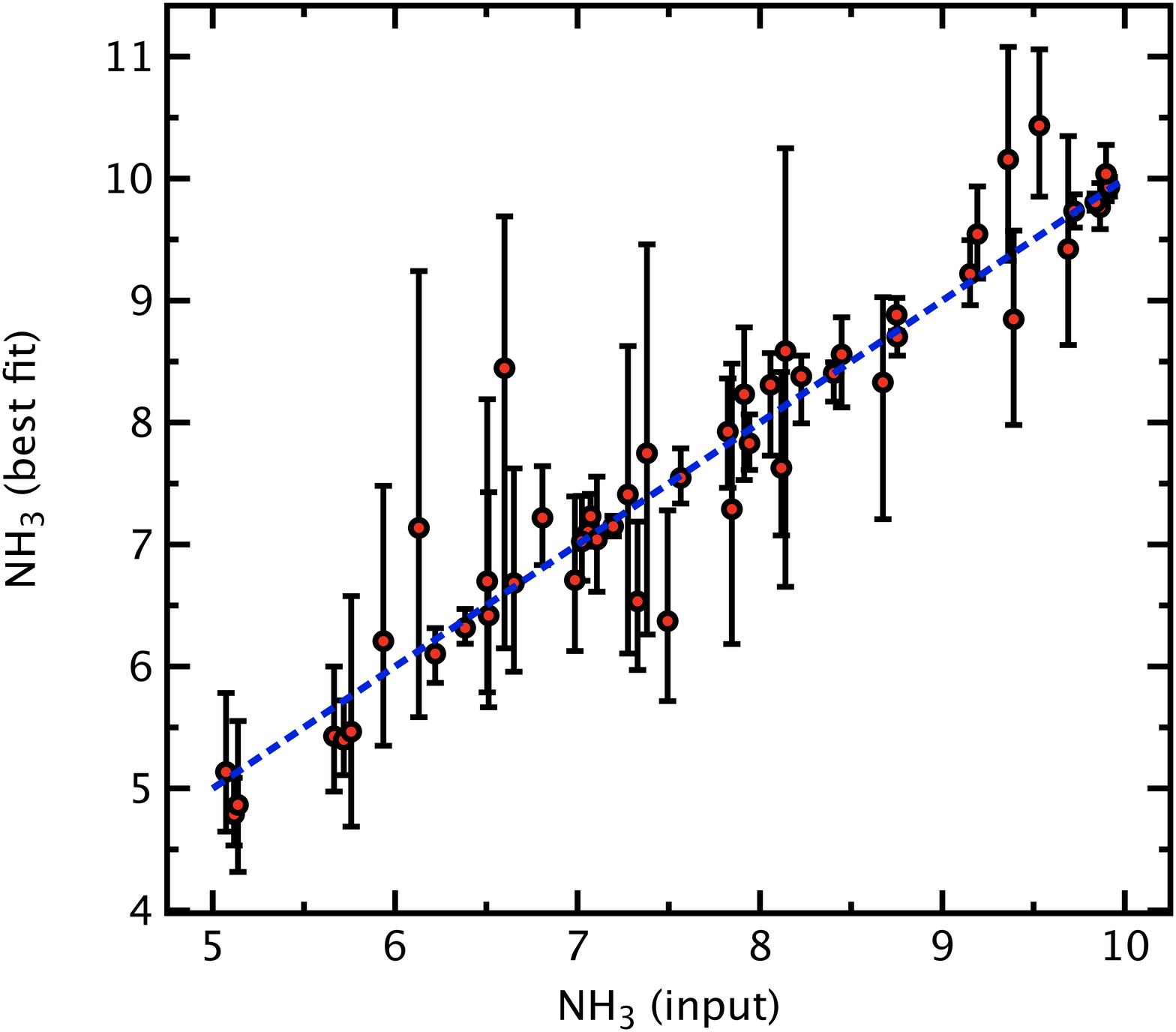} \includegraphics[width=0.33\hsize]{logxi3.eps}\includegraphics[width=0.33\hsize]{cvf3.eps}
 \caption{The best fit parameters versus the input parameters for the simulations of a 1 mCrab source observed for a 100 ks X-IFU observation (\rf=2, \afe=2, 1 mCrab, 100 ks, so-called configuration 3). 50 spins are generated with a constant spacing between 0 and 0.995. All other parameters are drawn from uniform distributions within bounds listed in Table \ref{table:1}. Top-left: the black hole spin, top-middle: the height of the coronal source and top-right: the photon index of its spectrum. Then downwards, from left-to-right: the parameters of the three absorbers: the \nh, the ionization (\logxi) and the covering factor. Errors are computed at the 90\% confidence level for variation of one single parameters. The mean errors for the spin and height of the irradiating source are $\sim 0.05$ and $\sim 0.18$ \rg\ respectively. The mean errors for all parameters are listed in Table \ref{table:2}. In particular, the power law index is accurately determined and shows no bias (see \S \ref{appendix_a}). }
         \label{conf3}
   \end{figure*}

\begin{figure*}
   \centering
   \includegraphics[width=0.45\hsize]{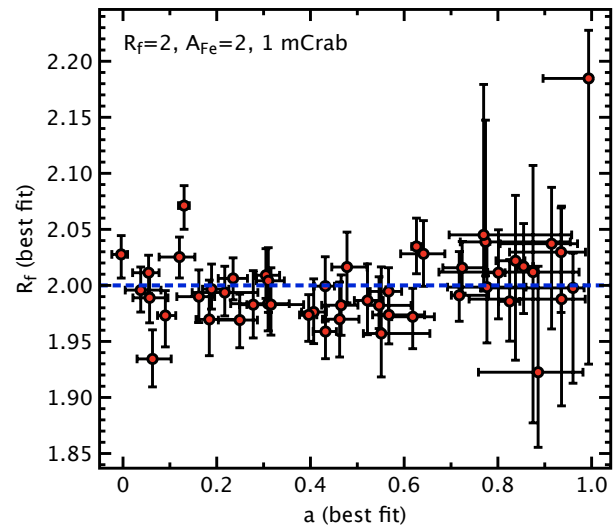}
    \caption{The reflection fraction recovered from the fit of configuration 3 spectra (\rf=2,\afe=2, 1 mCrab). This figure complements Fig. \ref{conf3}. }
         \label{conf3_ref}
   \end{figure*}

\begin{figure*}
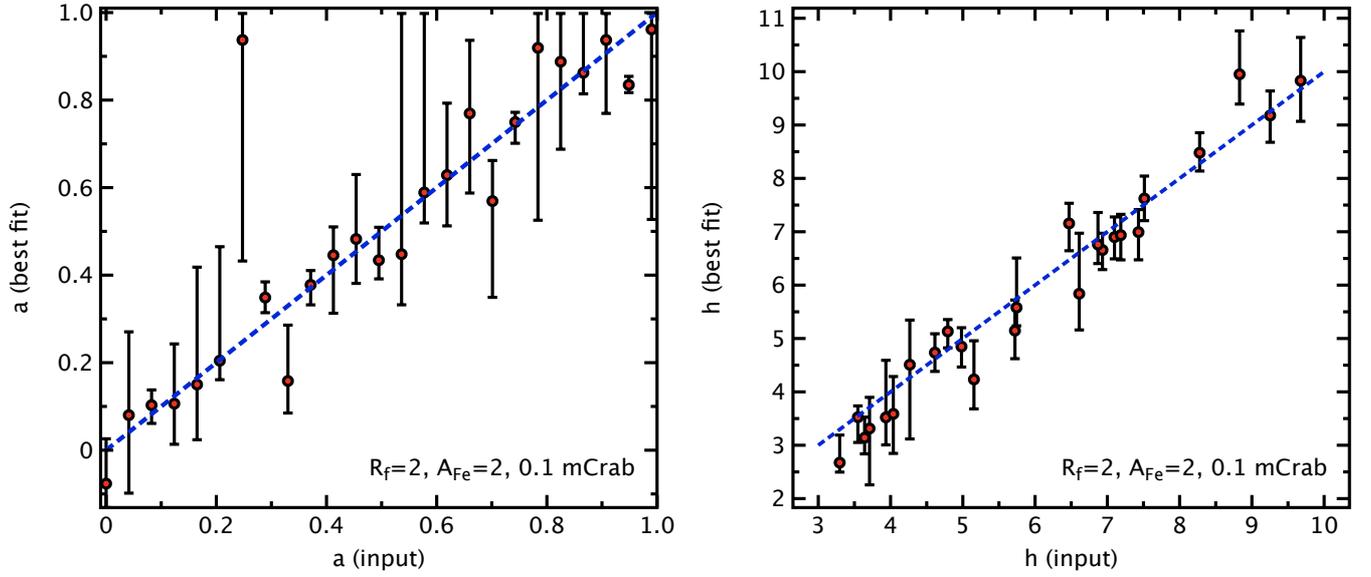

   \centering
      \includegraphics[width=0.49\hsize]{a.eps} \includegraphics[width=0.49\hsize]{h.eps}
     \caption{The best fit parameters versus the input parameters for the case of a 0.1 mCrab source observed for a 150 ks X-IFU observation (configuration 4, \rf=2, \afe=2). 50 spins are generated with a constant spacing between 0 and 0.995. The other parameters are allowed to vary with the range listed in Table \ref{table:1}. Left: the black hole spin, right: the height of the irradiating source against their input values. Errors are computed at the 90\% confidence level for variation of one single parameter. The mean errors on the spin and height of the irradiating source are $\sim 0.17$ and $\sim 0.6$ \rg\ respectively. The errors on all parameters are listed in Table \ref{table:2}.}
         \label{conf4}
   \end{figure*}

\begin{figure*}
   \centering
      \includegraphics[width=0.49\hsize]{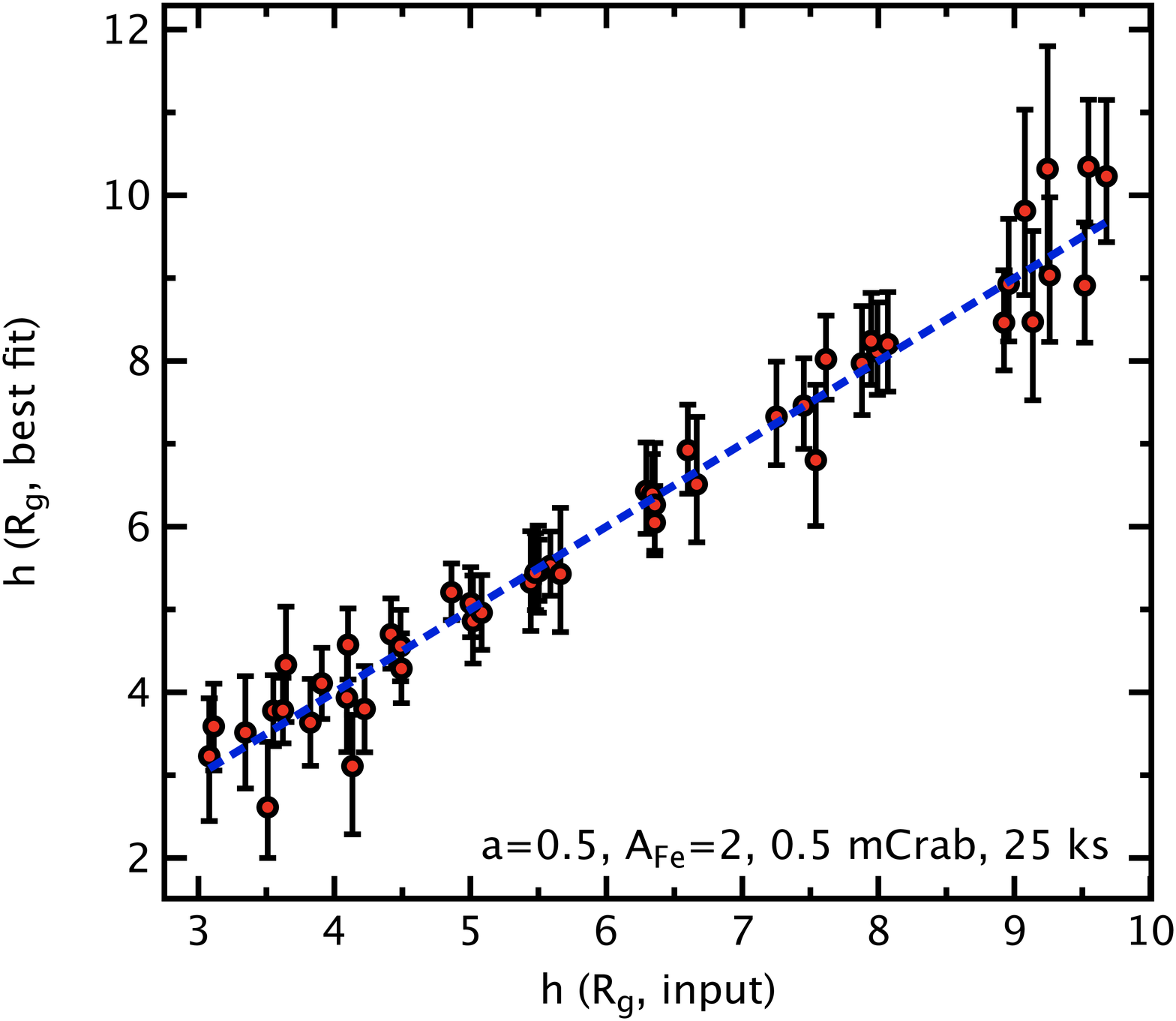} \includegraphics[width=0.49\hsize]{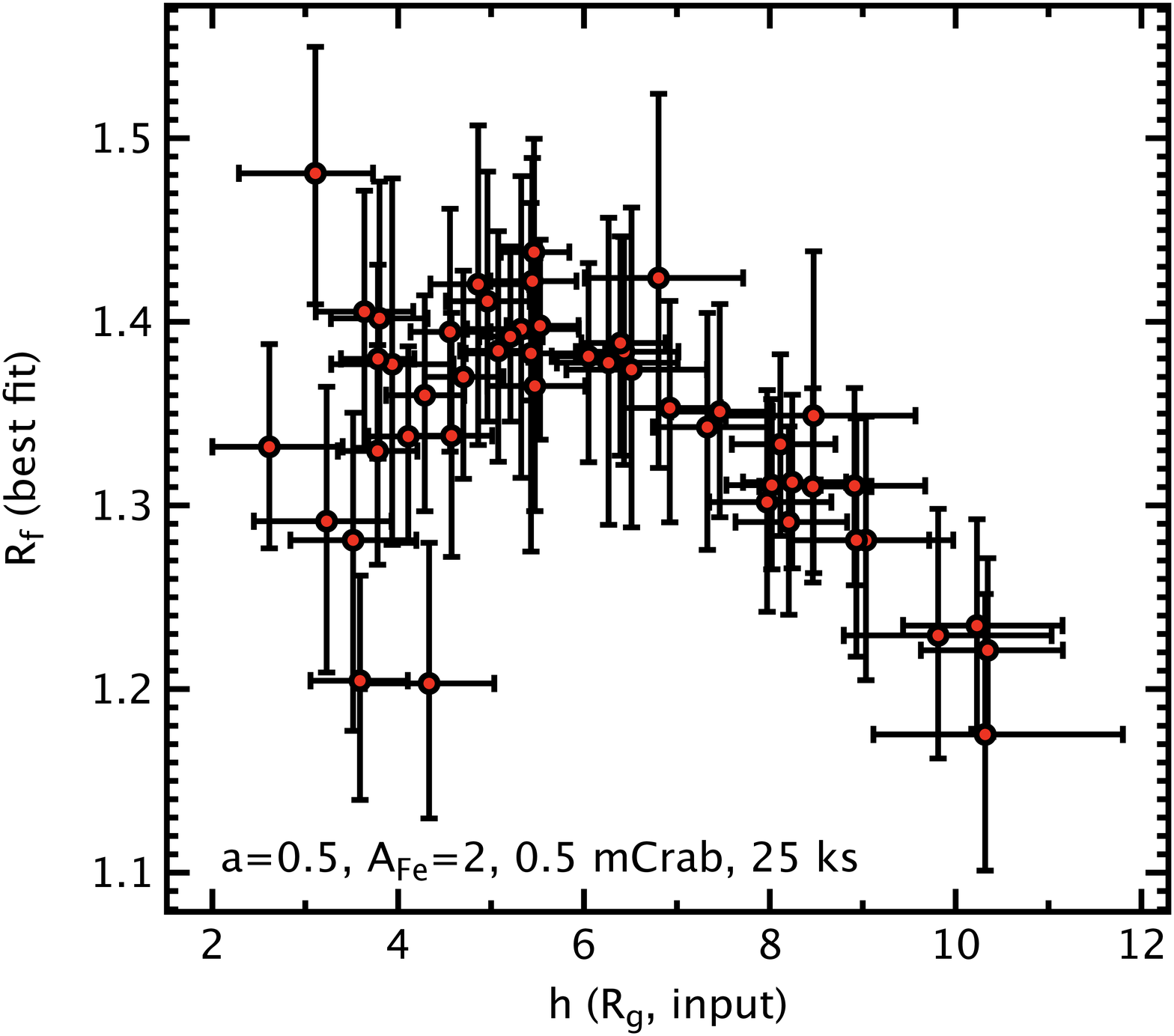} 
      \caption{The best fit parameters versus the input parameters for the case of a 0.5 mCrab source observed for a 25 ks X-IFU observation (configuration 5). The spin is fixed to 0.5 and the other parameters are allowed to vary with the range listed in Table \ref{table:1}. From left to right, the height of the coronal source against its input values, and the reflection fraction computed by the \relxill\ model against the height of the coronal source. Errors are computed at the 90\% confidence level for variation of one single parameters. As predicted by the model, \rf\ decreases with the height of the X-ray source. The mean error on $h$ is about 0.6 \rg.}
         \label{conf5}
   \end{figure*}

\begin{figure*}
   \centering
      \includegraphics[width=0.49\hsize]{cvf3.eps} \includegraphics[width=0.49\hsize]{logxi3.eps} 
      \includegraphics[width=0.49\hsize]{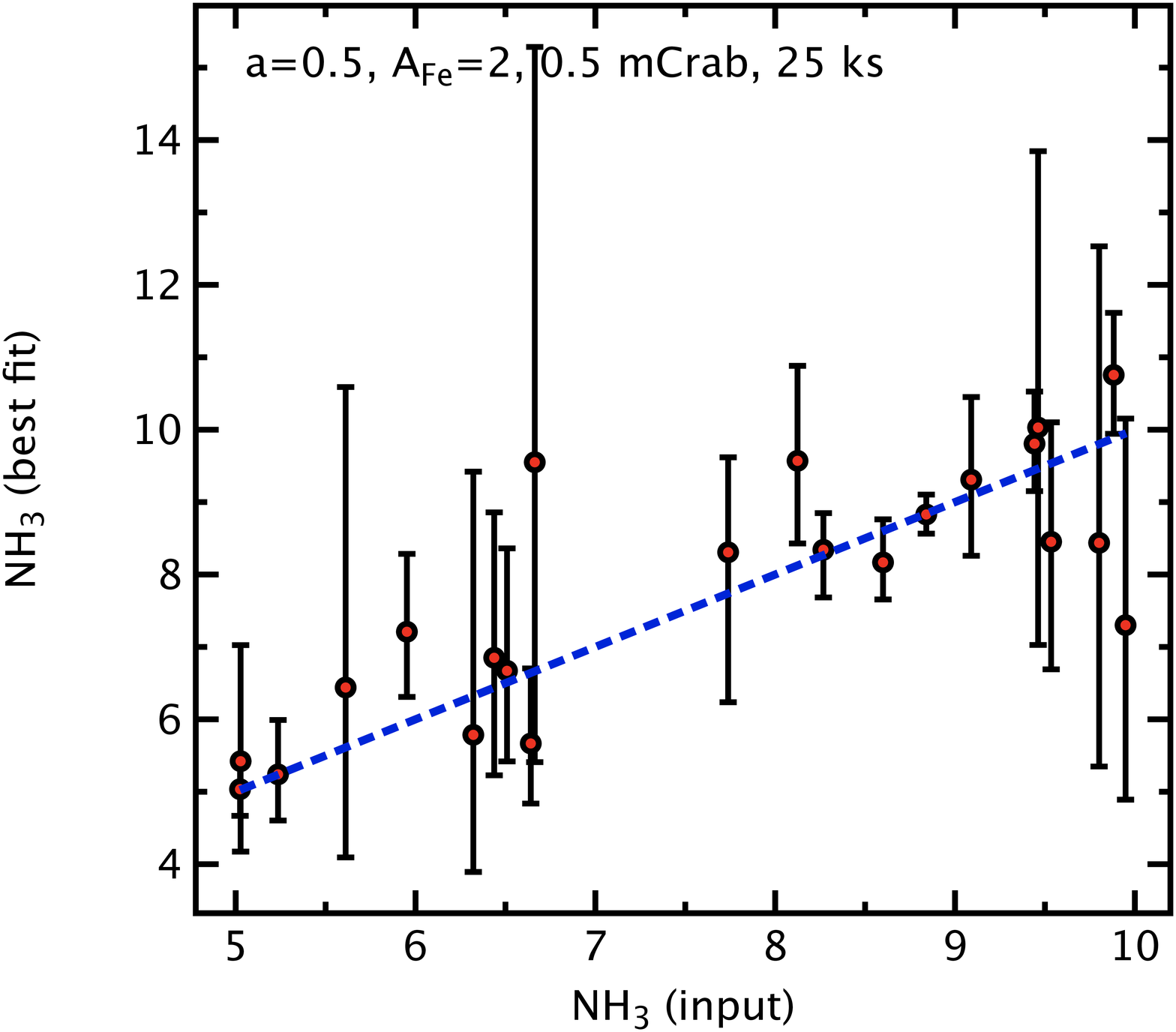} \includegraphics[width=0.49\hsize]{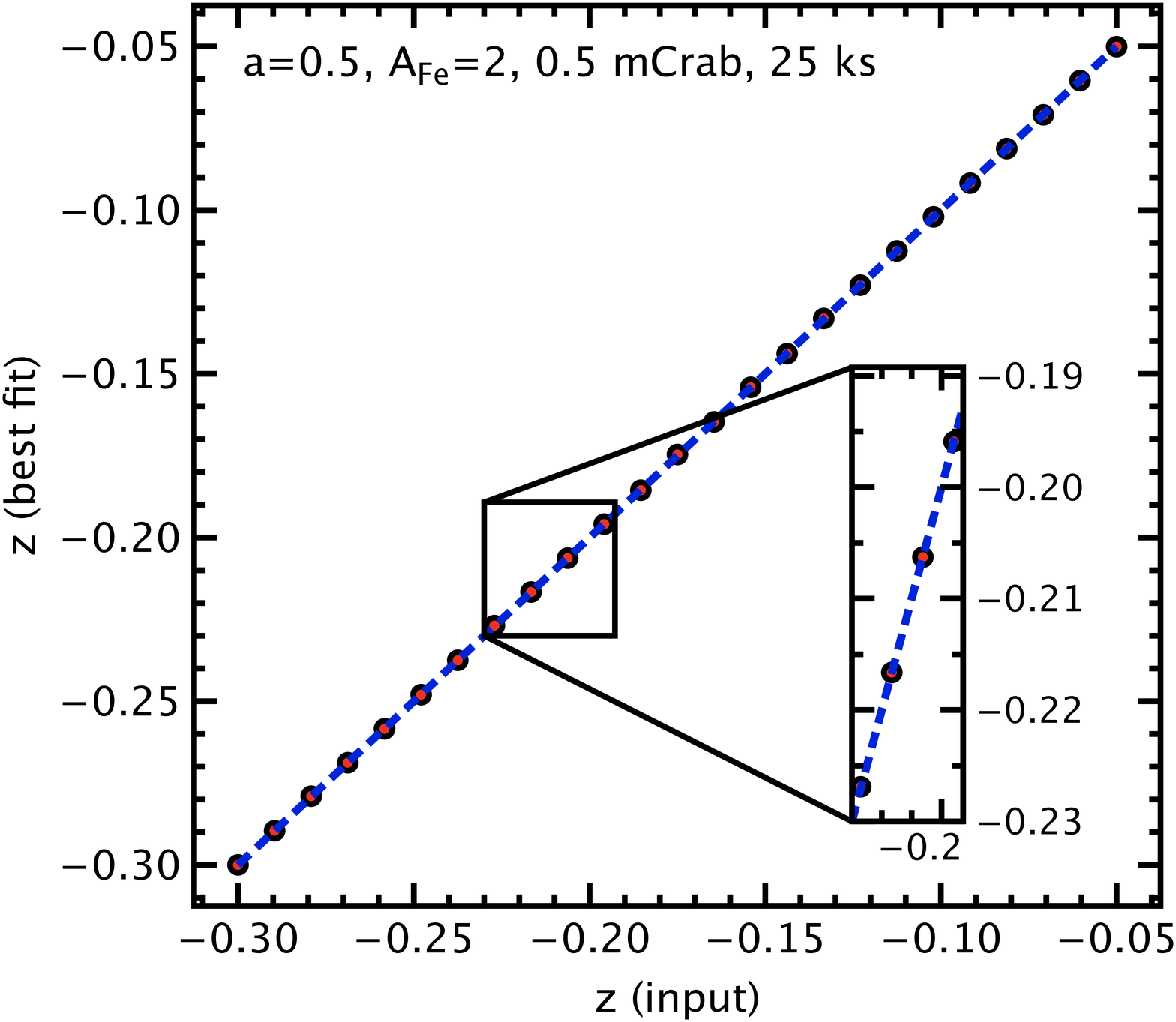} 
      \caption{The parameters of the high velocity absorber measured for an integration time of 25 ks for a 0.5 mCrab source (configuration 6). The absorber is smeared with a broadening velocity of 1000 km/s (at 6 keV) (see text for details). As can be seen, the redshift of the absorber is measured with very high accuracy.}
         \label{conf6}
   \end{figure*}

\begin{figure*}
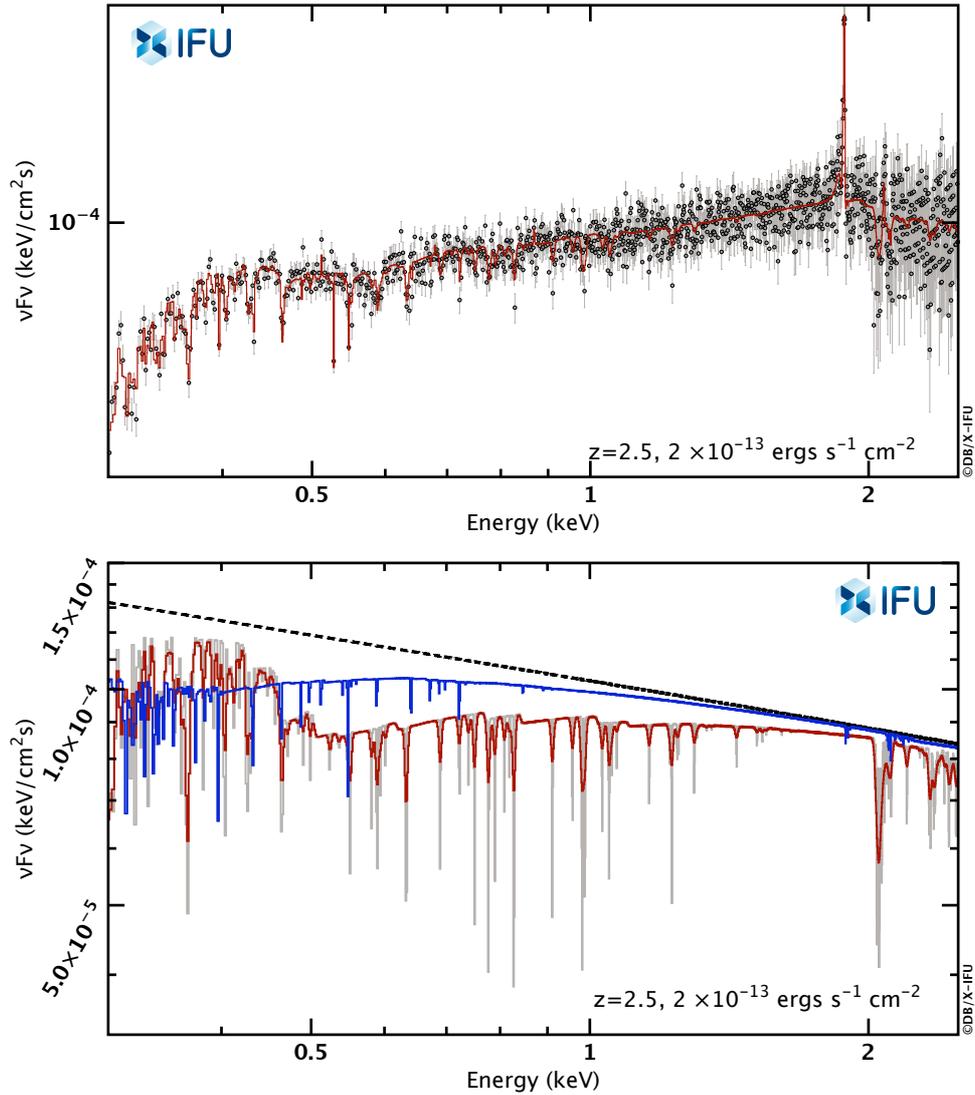

   \centering
      \includegraphics[width=0.70\hsize]{eeufspec.eps} \includegraphics[width=0.70\hsize]{powerlaw_wo_ref.eps} 
      \caption{Top: The energy spectrum of a high-redshift AGN (z=2.5) observed with X-IFU for 100 ks. Bottom: The incident power law spectrum (reflection component removed, black dashed line) to highlight the imprints of the absorbers. The low density, low ionization absorber is shown with the blue curve (\nh= $0.9~10^{22}$ cm$^{-2}$, \logxi=1.8) at a redshift of 2.46. The high density high velocity absorber (\nh= $7.1~10^{22}$ cm$^{-2}$, \logxi=2.9) whose redshift is 2.33 is shown without smearing in grey.  A velocity broadening normalized to 1000 km/s at 6 keV (rest frame) has been assumed and smears out the absorber features (red line). The black hole spin has been assumed to be 0.5. The iron abundance has been set to 1. Fitting such spectrum would enable the redshift to the source and the velocity of the outflow to be measured with an extremely high accuracy of $\sim 0.001$ and $\sim 0.01$ respectively (statistical error only). This is due to the strong narrow iron line produced by the distant reflector and the large number of absorption lines due the two absorbers.}
         \label{fig_distant_agn}
   \end{figure*}

\begin{figure*}
   \centering
      \includegraphics[width=0.55\hsize]{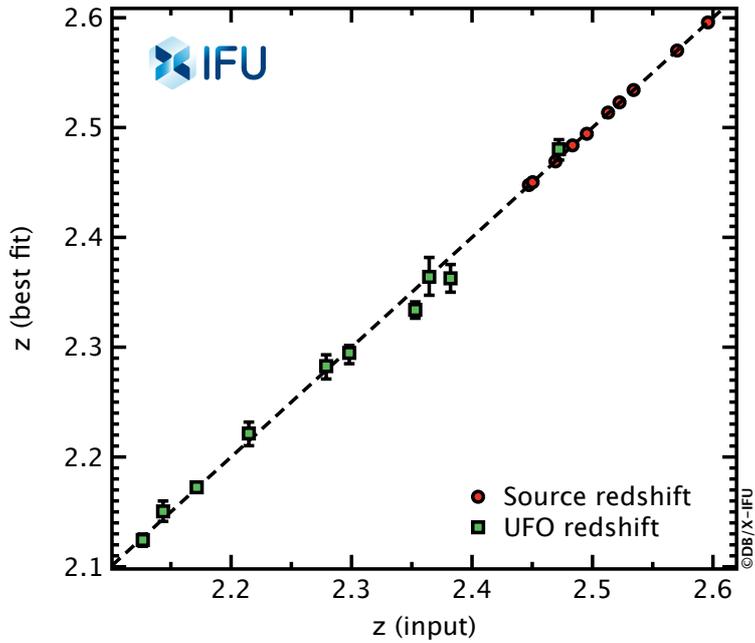} 
            \caption{The redshift of the source (filled red circles) and the redshift of the high velocity blue shifted absorber (filled green symbols), against their input values in the simulations.  10 source redshifts are drawn from a uniform distribution between 2.4 and 2.6, while 10 blue shifts are added from a uniform distribution bounded between -0.3 and -0.05. A large velocity broadening normalized to 3000 km/s has been assumed for the high velocity absorber. The spin of the source is assumed to be 0.5. For the reflection parameters, we assumed \rf=1, \afe=1. The flux of the source is $2 \times 10^{-13}$ \ergs, and the integration of the spectrum is 100 ks. See \S\ref{ufo} for details. The 90\% confidence level errors are plotted. The errors are less than 0.001 and 0.01 for the source redshift and UFO velocity respectively. }
         \label{fig_redshifts}
   \end{figure*}

\begin{figure*}[!t]
   \centering{\includegraphics[width=0.95\hsize]{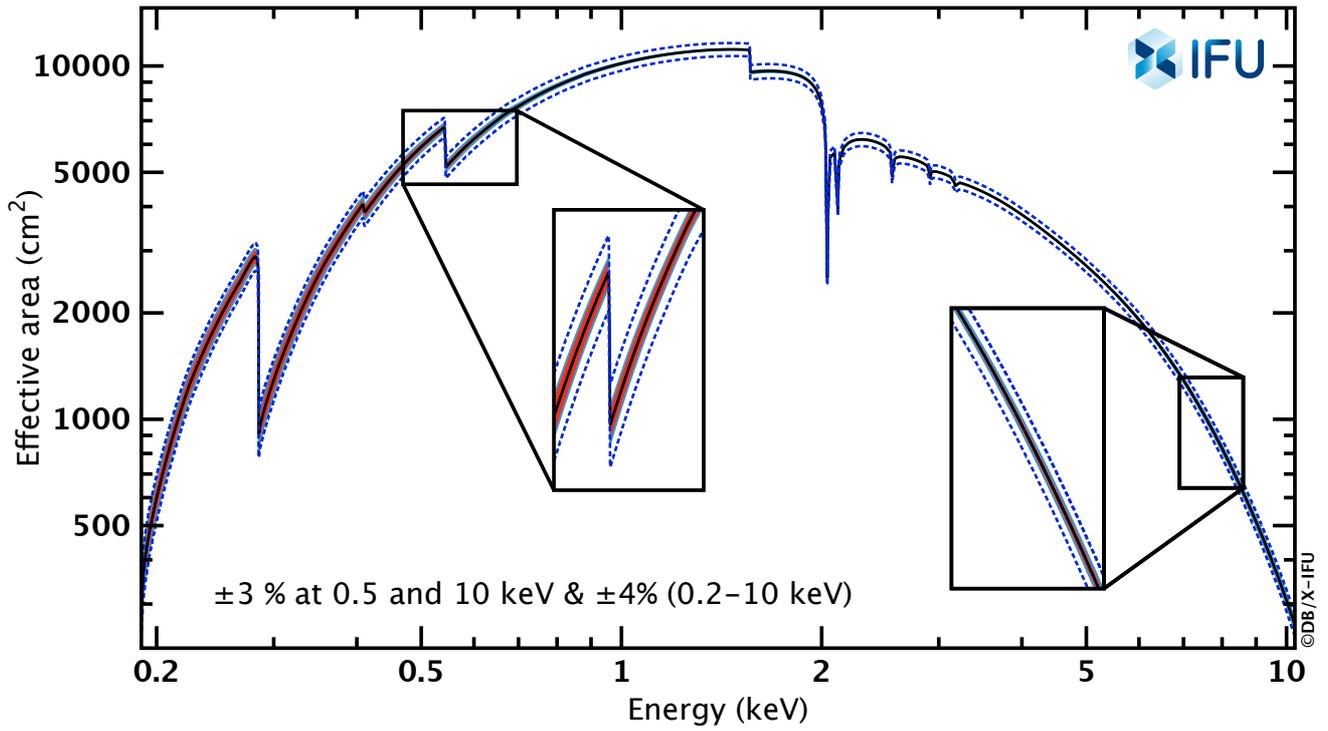}}
    \caption{The envelope of the X-IFU responses considered to assess the potential systematics related to calibration uncertainties. The 3\% ($1 \sigma$) on the relative broad band effective area is shown in filled red around the nominal response which is indicated with a solid black line. The 4\% (1$\sigma$) on the absolute effective area knowledge is delimited by the blue dashed line and comes on top of the previous one. 1000 response files are drawn from within the envelope, assuming non truncated normal distributions. See \S\ref{calib} for details about the method.} 
    \label{fig_aeff}
   \end{figure*}

\begin{figure*}
 \centering{
    \includegraphics[width=0.45\hsize]{a.eps}\includegraphics[width=0.45\hsize]{h.eps}
     \includegraphics[width=0.45\hsize]{gamma.eps}\includegraphics[width=0.45\hsize]{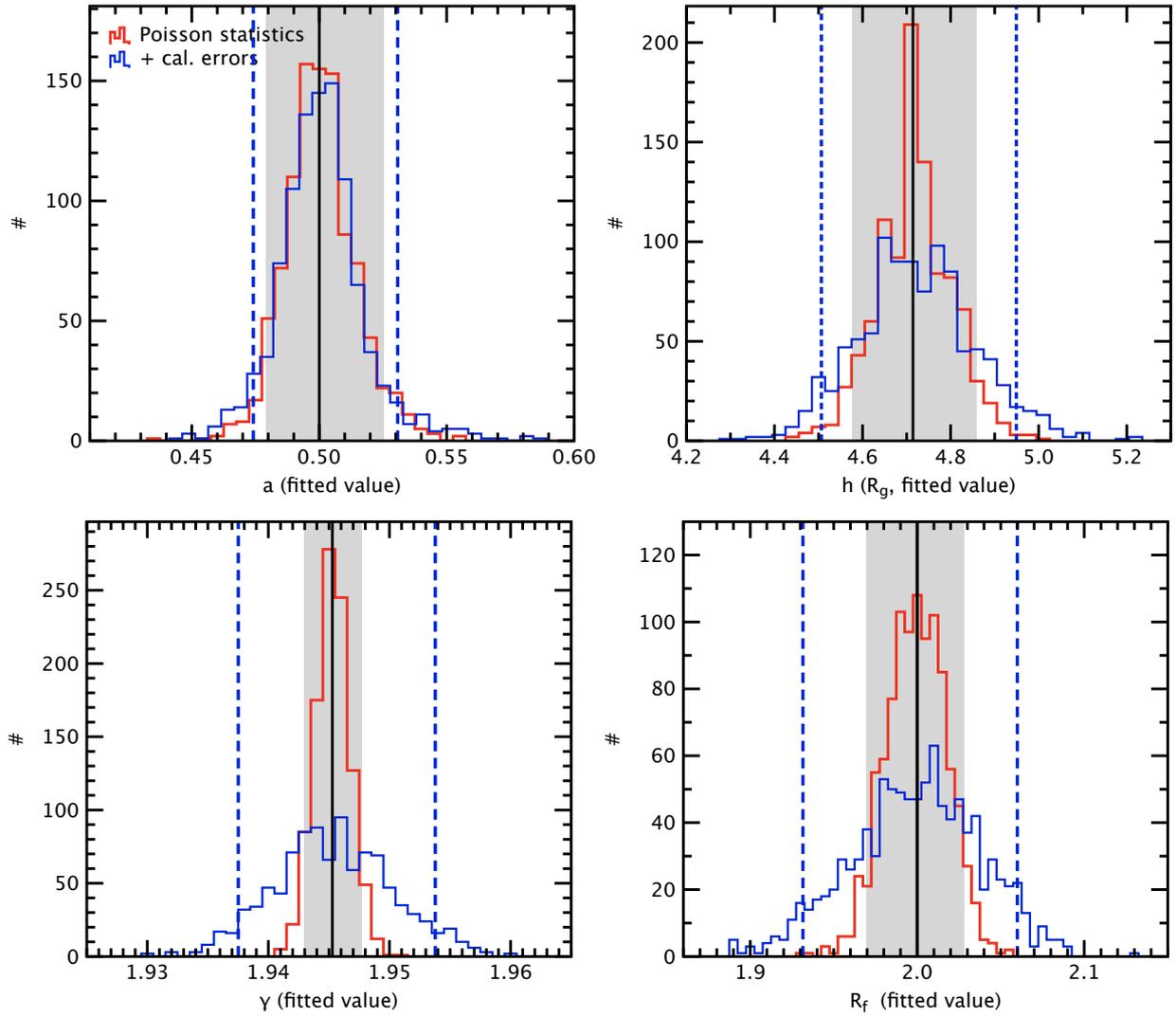}
   }
    \caption{The histogram distribution of best fit parameters arising from Poisson statistics alone (in red, the 90\% quantile is indicated with the grey area) and from Poisson statistics and calibration errors (in blue, the 90\% quantile is delimited with the blue dashed line). Top-left: the spin parameter, Top-right: the height of the irradiating source, bottom-left: the power law index, and bottom-right: the reflection fraction. The case simulated here corresponds to a spin of 0.5, a 1 mCrab source observed for 100 ks, with a reflection fraction set to 2. As can be seen, at the level of calibration errors considered here, the spin parameter and the height of the X-ray source do not suffer from any significant systematics, while the systematic errors on the power law index and reflection fraction are larger, but remain small. No biases are introduced in all cases. See \S\ref{calib}.} 
    \label{fig_res_cal_uncertainties}
   \end{figure*}

\begin{figure*}
 \centering{\includegraphics[width=0.65\hsize]{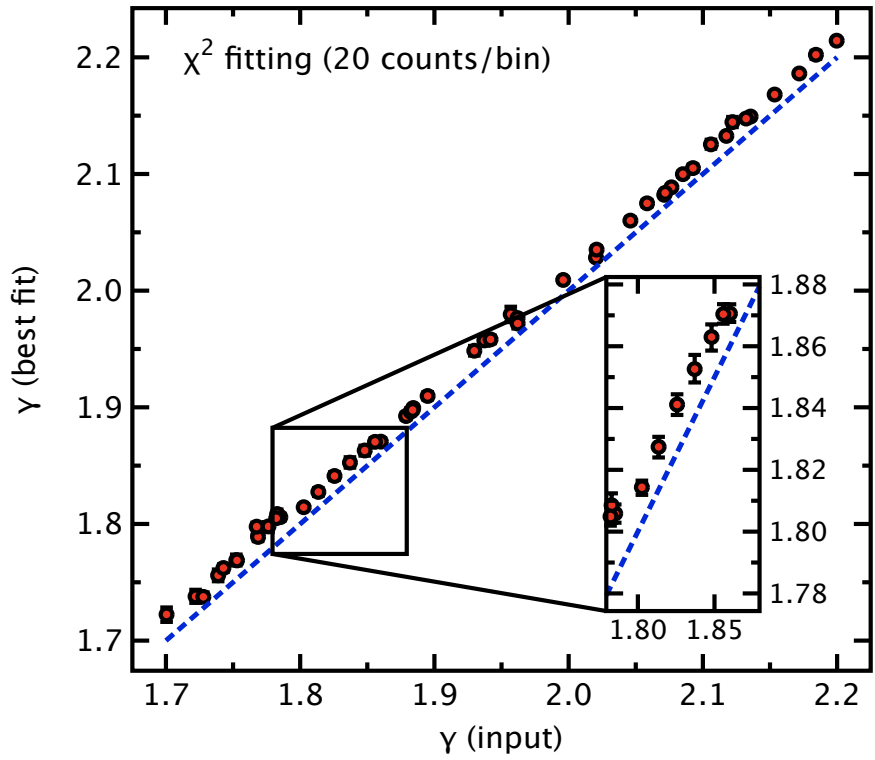}}
    \caption{The best fit power law index with its 90\% confidence level error is plotted against the input value to show that with $\chi^2$ fitting a bias larger than the statistical error is introduced. The input power law index of the irradiating source is drawn uniformly between 1.7 and 2.2 (50 values). Before fitting, the spectra are binned to have a minimum number of 20 counts per bin. No such bias is seen when the \cstat\ is used, as indicated in Figure \ref{conf3} (top-right panel). } 
    \label{fig_bias_pl_index}
   \end{figure*}

$\chi^2$ statistics is often used as a the fitting metric, although its limitations are known, especially in the low count regime. As discussed by \citet{humphrey_2009_apj}, even in the high count rate regime (when the counts per bin gets typically larger than $\sim 20$), $\chi^2$ fitting will lead to biased parameter estimates, unless the number of data bins is far smaller than the square root of the number of counts in the spectrum (which is not the case for most simulations presented here). The bias may be comparable to, or even exceed, the statistical error. We have repeated the configuration 1 simulation, replacing the optimal binning scheme of \citet{kaastra_2016} by a standard grouping scheme ensuring that each spectral bin would have at least 20 counts. We have used $\chi^2$ statistics. Of all the 16 free parameters of the fit, the photon index of the power law has a very small statistical error (0.003 in table \ref{table:2}). In figure \ref{fig_bias_pl_index}, the best fit power law index is reported against the input power law index. As can be seen, a bias is present towards recovering steeper indexes, and the bias exceeds the statistical error. The bias is still present when the data are grouped further having a minimum of 50 counts per bin. A similar bias was present in the simulations reported by \cite{choudhury_2017}. No such bias is present in our fits based on \cstat, as shown in Figure \ref{conf3}. To conclude, for X-IFU data, it is recommended to always use \cstat\ in fitting spectra, see also \cite{kaastra_2017_aa} on how \cstat\ can be used for statistical tests, such as assessing the goodness of fit of a spectral model, as used here.

\begin{acknowledgements}
The authors wish to thank the anonymous referee for useful comments.

DB acknowledges useful discussions with Thomas Dauser, Edoardo Cucchetti, Chris Done, Jeremy Drake and Etienne Pointecouteau. DB and MC thank Laura Brenneman, Francisco Carrera, Mauro Dadina, Javier Garcia, Matteo Guainazzi, Jelle Kaastra, Elias Kammoun, Giovanni Miniutti, Jon Miller, and Jiri Svoboda for their useful suggestions and comments on an earlier version of the paper. Special thanks to Edoardo Cucchetti and the computer support team at IRAP (Elodie Bourrec and C\'edric Hillembrand) for providing DB with the cluster resources required to carry out these time consuming simulations. MC acknowledges financial support from the Italian Space Agency under agreement ASI-INAF n.2017-14-H.O.and 2018-11-HH.0. DB acknowledges support from the French Space Agency (CNES). DB wishes to dedicate this paper to his beloved mother who passed away along the preparation of this work.
 \end{acknowledgements}

%
%

\graphicspath{{figures/fig1/}}
\clearpage  
\graphicspath{{figures/conf2/}}
 \clearpage   
\graphicspath{{figures/conf3/}}
 \clearpage  

\clearpage     
\graphicspath{{figures/conf4/}}
  \clearpage   
   \graphicspath{{figures/conf5/}}
 \clearpage    
    \graphicspath{{figures/conf6/}}
\clearpage  
    \graphicspath{{figures/fig8/}}
\clearpage     
      \graphicspath{{figures/redshift_highz/}}
 
\clearpage   
   \graphicspath{{figures/fig9/}}
   
\clearpage  \graphicspath{{figures/fig10/}}
  
\clearpage  
\graphicspath{{figures/fig11/}}

\end{document}